\documentclass[10pt,prd,aps,showpacs,eqsecnum,amsfonts,amsmath,superscriptaddress,preprintnumbers,a4paper,twocolumn]{revtex4-1}

\usepackage{color}
\usepackage{psfrag}
\usepackage{graphicx}
\usepackage{multirow}
\usepackage{amssymb,amsfonts,latexsym}
\usepackage{subfigure}  
\usepackage{bm}

\usepackage{url}

\makeatletter
\def\url@leostyle{
  \@ifundefined{selectfont}{\def\UrlFont{\sf}}{\def\UrlFont{\small\ttfamily}}}
\makeatother
\begin{document}

\newcommand{\UIB}{Departament de F\'isica, Universitat de les Illes
  Balears, Crta. Valldemossa km 7.5, E-07122 Palma, Spain}
\newcommand{\AEIGolm}{Max-Planck-Institut f\"ur Gravitationsphysik
  (Albert-Einstein-Institut), Am~M\"uhlenberg 1, 14476~Golm, Germany}
\newcommand{\Jena}{Theoretisch-Physikalisches Institut, Friedrich
  Schiller Universit\"at Jena, Max-Wien-Platz 1, 07743~Jena, Germany}
\newcommand{\LIGOCaltech}{LIGO Laboratory, California Institute of Technology, 
Pasadena, CA 91125, U.S.A.}
\newcommand{\TAPIR}{Theoretical Astrophysics, California Institute of
  Technology, Pasadena, CA 91125, U.S.A.}
\newcommand{\Vienna}{Faculty of Physics,
  University of Vienna, Boltzmanngasse 5, A-1090 Vienna, Austria}
\newcommand{\NASAGoddard}{NASA Goddard Space Flight Center, Greenbelt,
  MD 20771, USA} 

\title{Matching post-Newtonian
  and numerical relativity waveforms: systematic errors and a new
  phenomenological model for non-precessing black hole binaries}

\author{L.~Santamar\'ia}
\affiliation{\AEIGolm}

\author{F.~Ohme}
\affiliation{\AEIGolm}

\author{P.~Ajith}
\affiliation{\LIGOCaltech}
\affiliation{\TAPIR}

\author{B.~Br\"ugmann}
\affiliation{\Jena}

\author{N.~Dorband}
\affiliation{\AEIGolm}

\author{M.~Hannam}
\affiliation{\Vienna}

\author{S.~Husa}
\affiliation{\UIB}

\author{P.~M\"osta}
\affiliation{\AEIGolm}

\author{D.~Pollney}
\affiliation{\UIB}

\author{C.~Reisswig}
\affiliation{\TAPIR}

\author{E.~L.~Robinson}
\affiliation{\AEIGolm}

\author{J.~Seiler}
\affiliation{\NASAGoddard}

\author{B.~Krishnan}
\affiliation{\AEIGolm}


\bigskip

\date{\today}

\begin{abstract}
  We present a new phenomenological gravitational waveform model for
  the inspiral and coalescence of non-precessing spinning black hole
  binaries. Our approach is based on a frequency domain matching of
  post-Newtonian inspiral waveforms with numerical relativity based
  binary black hole coalescence waveforms.  We quantify the various
  possible sources of systematic errors that arise in matching
  post-Newtonian and numerical relativity waveforms, and we use a
  matching criteria based on minimizing these errors; we find that the
  dominant source of errors are those in the post-Newtonian waveforms
  near the merger. An analytical formula for the dominant mode of the
  gravitational radiation of non-precessing black hole binaries is
  presented that captures the phenomenology of the hybrid
  waveforms. Its implementation in the current searches for
  gravitational waves should allow cross-checks of other
  inspiral-merger-ringdown waveform families and improve the reach of
  gravitational wave searches.
\end{abstract}


\preprint{LIGO-P1000048-v3}
\preprint{AEI-2010-122}

\maketitle

\section{Introduction}
\label{sec:intro}

As a generalization of the classic Kepler problem in Newtonian
gravity, the binary black hole (BBH) system in general relativity is
of great interest from a fundamental physics viewpoint.  Equally
importantly, this system has received a great deal of attention for
its relevance in astrophysics and, in particular, as one of the most
promising sources of detectable gravitational radiation for the
present and future generations of gravitational-wave detectors, such as
LIGO~\cite{Abbott:2007kv}, Virgo~\cite{Acernese:2008zzf},
GEO600~\cite{Grote:2008zz}, LISA~\cite{Shaddock:2009za} or the
Einstein Telescope~\cite{Punturo10}.  The Kepler problem can be solved
exactly in Newtonian gravity and it leads to the well-known elliptical
orbits when the system is gravitationally bound.  In contrast, in
general relativity, closed orbits do not exist and the BBH system
emits gravitational waves (GWs) which carry away energy, thereby
causing the black holes to inspiral inwards, and to eventually
coalesce.  The emitted GWs are expected to carry important information
about this process, and it is one of the goals of gravitational wave
astronomy to detect these signals and decode them.

No analytic solutions of Einstein's equations of general relativity
are known for the full inspiral and merger of two black holes.
Post-Newtonian (PN) methods can be used to calculate an accurate
approximation to the early inspiral phase, using an expansion in
powers of $v/c$ (where $v$ is the orbital velocity and $c$ is the
speed of light).  As for the coalescence phase, starting with
\cite{Pretorius:2005gq,Campanelli:2005dd,Baker05a}, the late inspiral
and merger has been calculated by large-scale numerical solutions of
the full Einstein field equations. Since the initial breakthroughs in 2005,
there has been dramatic progress in numerical relativity (NR)
simulations for GW astronomy, including many more orbits before
merger, greater accuracy and a growing sampling of the
black hole-binary parameter space.  A summary of the published
``long'' waveforms is given in the review~\cite{Hannam:2009rd}, and a
complete catalog of waveforms is being compiled at~\cite{ninja-web};
more recent work is summarized in~\cite{Hinder:2010vn}.  NR results
are now accurate enough for GW astronomy applications over the next
few years~\cite{Hannam:2009hh}, and have started playing a role in
GW searches \cite{Aylott:2009ya,Aylott:2009tn}.

Given PN and NR results, it is promising to try and combine them to
produce ``complete'' inspiral-merger-ringdown waveforms. PN techniques
in their standard formulation become less accurate as the binary
shrinks, and the approximation breaks down completely somewhere prior
to the merger.  NR waveforms, on the other hand, become more and more
computationally expensive the larger the number of cycles that one
wishes to simulate; the longest published data spans 16 orbits for
the equal-mass 
non-spinning case~\cite{Scheel:2008rj}.  We therefore would hope to
combine PN and NR results in the region between the point where NR
simulations start and where PN breaks down. To do this it is critical
to verify that the PN and NR results are in good agreement in this
region and that there is a consistent PN-NR matching procedure. Much
work has been done in comparing PN and NR results over the last 5-15
orbits before merger for a variety of physical configurations, such as
the equal-mass non-spinning 
case~\cite{Buonanno:2006ui,Baker:2006ha,Hannam:2007ik,Gopakumar:2007vh,Boyle:2007ft,Boyle:2008ge,Hinder:2008kv}, 
the equal-mass non-precessing-spin case~\cite{Hannam:2007wf},
and the unequal-mass spinning case~\cite{Campanelli:2008nk}. The
consistency of PN amplitudes during the merger and ringdown has
also been studied~\cite{Buonanno:2006ui,Berti:2007fi,Berti:2007nw}. These
studies suggest that a sufficiently accurate combination of PN and NR
results should be possible. One topic that has not received much
attention, however, is the systematic errors that are introduced by
different choices of matching procedure.

One of the aims of this paper is to further understand and quantify
the various systematic errors that arise in the matching procedure.
There are thus far two kinds of approaches to the PN-NR matching
problem, both of which have yielded successful results.  The first is
the Effective-One-Body (EOB) approach
\cite{Buonanno:1998gg,Buonanno:2000ef,Damour:1997ub,Damour:2000we}.
Originally motivated by similar techniques in quantum field theory,
the idea is to map the two body problem into an effective one-body
system with an appropriate potential and with the same energy levels
as the two-body system.  It was shown \cite{Buonanno:1998gg} that the
appropriate one-body problem (for non-spinning black holes) is that of
a single particle moving in a deformed Schwarzschild spacetime.  It
turns out that most parameters of this one-body system can be found by
using the appropriate PN calculations, and the remaining parameters
are calculated by calibrating to NR
simulations. This approach has been successful so far for non-spinning
systems where only a few parameters need to be calibrated by
NR~\cite{Buonanno:2007pf,Buonanno:2009qa,Damour:2007vq,Damour:2008te,Damour:2009kr}. 
The spinning case is more complicated, and work is underway to extend
the parameter space described by the model~\cite{Pan:2009wj}.

A complementary approach is to perform a phenomenological matching of
the GW waveforms in a window (which could be either in the time or
frequency domain) where both PN and NR are expected to be good
approximations to the true waveform.  The first step is to construct a
hybrid PN-NR waveform by matching the two waveforms within the
matching window.  The waveform is completely PN before this window,
completely NR afterward, and it interpolates between the two in the
matching window.  Once the hybrid waveform is constructed and we are
confident about the matching procedure, the second step is to fit the
hybrid waveform to a parametrized model containing a number of
phenomenological coefficients and finally to map them to the physical
parameters of the system.  The resulting model would thus be
parametrized by the masses and spins of the two black holes (and
eccentricity if appropriate).  Most of the work in this approach has
thus far been based on matching PN and NR waveforms in the time
domain, but then producing a phenomenological model in the frequency
domain, which is often more convenient for data-analysis
applications~\cite{Ajith:2007qp,Ajith:2007xh,Ajith:2007kx,Ajith:2009bn}.
See also \cite{Sturani:2010yv} for a complementary construction.  
In this paper we take a slightly different approach: both the
construction of the PN-NR hybrid waveform and the matching to a
phenomenological model are carried out in the frequency domain.  The
reasons for this are twofold.  First, we find it easier to
work in the frequency domain since the quantities used to estimate the
errors of our matching procedure and the goodness of the fit, such as
waveform overlaps, are conveniently formulated in Fourier space.
Second, and more importantly, in light of the potential errors in the
hybrid construction, comparing results between two independent methods
is a valuable way of ensuring that the matching procedure is robust.
The frequency domain construction presented here is complementary to
the time domain method of~\cite{Ajith:2009bn}.

The phenomenological waveform family presented in~\cite{Ajith:2009bn}
used a simple piecewise ansatz for the phase of the hybrid PN-NR
waveform, and another for the amplitude. The resulting analytic model
was found to agree with the hybrid waveforms with overlaps above 97\%
for most black hole binary systems that would be observable by the
current LIGO detectors.  In this paper we investigate whether the
fidelity of the phenomenological waveforms can be improved by using
ans\"atze that make smooth transitions between their inspiral, merger
and rindown forms. This procedure also allows us to further test the
robustness of the phenomenological model's construction to variations
in its analytic form.

The main results of this paper are the following. We construct hybrid
waveforms for binary black hole systems with aligned spins in the
frequency domain. We do this by combining 3.5PN waveforms in the
stationary phase approximation with a number of NR results.  We show
that this construction is internally consistent and it yields hybrids
which are, for the most part, sufficiently accurate for the initial
and advanced LIGO detectors.  Notably, the difference between the
different PN approximants is a more significant source of error than
the numerical errors in the NR waveforms.  Using these hybrid
waveforms, we construct a phenomenological frequency-domain waveform
model depending on three parameters (as in \cite{Ajith:2009bn}) and
covering the space of aligned spins and moderate mass ratios.  We show
that the model fits the original hybrid waveforms with the overlaps
(maximized over the model parameters) better than 97\% for Advanced
LIGO (and for the most part, better than 99\%) for essentially all
black hole systems observable with Advanced LIGO, i.e. for systems
with total mass ranging up to $\sim 400 M_\odot$. These results are
comparable to those obtained in~\cite{Ajith:2009bn}, suggesting that
the phenomenological construction is indeed robust and that the
phenomenological model waveforms are useful for detection purposes.  

Sections~\ref{sec:numrel} and~\ref{sec:postnewton} describe the
post-Newtonian waveform model and the numerical waveforms that we
employ.  Section~\ref{sec:pn-vs-nr} describes the fitting procedure and 
the various systematic errors that appear in this procedure. It
quantifies the reliability of the waveforms for specific GW detector
and signal-to-noise ratios (SNRs). Section~\ref{sec:fitting} fits these
hybrid waveforms to an analytic model. It shows that the model
provides a good representation of the hybrid waveforms and can be used
in GW searches in the appropriate parameter space.  
Finally, section~\ref{sec:summary} 
concludes with a summary and suggestions for future work.

\section{Numerical simulations of non-precessing black hole binaries}
\label{sec:numrel}

In this section we summarize the numerical waveforms used in this
paper.  Since the first successful numerical simulations of
equal-mass, non-spinning binary black hole mergers were
published~\cite{Pretorius:2005gq,Campanelli:2005dd,Baker05a} the NR
community has continued exploring the parameter space of the BBH
system. 
Each black hole is described by a mass and a spin
vector, and the binary's trajectory is described by adiabatically 
evolving Keplerian orbits, so 17 parameters are needed to describe the 
binary system (see e.g. \cite{Apostolatos:1994mx}).  Besides the two masses and 
spin vectors, we need a fiducial time $t_0$ and orbital phase $\phi_0$
at $t_0$, the distance to the source and its sky-location, two
parameters for the unit vector normal to the orbital plane, and
finally, if non-circular orbits are considered, we additionally need the
eccentricity and the direction of the semi-major axis.  Sufficiently
close to or during the merger, this description in terms of Keplerian
orbits will break down, and higher order black hole multipoles might
play a role as well.  

\begin{table*}
        \caption{NR codes and configurations used for the construction
          and verification of our hybrid waveforms and
          phenomenological model. The mass ratio $q$ is defined as
          $m_1/m_2$, assuming $m_1 \geqslant m_2$; $\chi_{1,2}$ are
          the dimensionless spins defined in Eq.~\eqref{eq:spin}; a
          positive value of $\chi_{1,2}$ means that the spin is
          aligned with the orbital angular momentum $\bm{L}$, and
          negative values are anti-aligned.}
        \label{tab:NRWaveforms}
    \begin{center}
        \begin{tabular*}{\linewidth}{@{\extracolsep{\fill}}c c c c c}
            \hline
            \hline

Data Set & Code & Mass ratios & Spins & Extraction of GW signal \\
\hline

\#1 &\texttt{BAM}~\cite{Brugmann:2008zz,Husa:2007rh} & $q \in
\{1,1.5,2,2.5,3,3.5,4\}$ & $(\chi_1,\chi_2) =(0,0)$ & at $r=90 M$   \\

\#2 &''& $q =1$ &
$(\chi_1,\chi_2) = (a,a)$, $a \in \pm \{0.25,0.5,0.75,0.85\}$ & '' \\

\#3 &''& $q \in \{2,3,4\}$ &
$(\chi_1,\chi_2) = (a,a)$, $a \in \{\pm 0.5,0.75\}$ & ''  \\

\#4 &'' & $q =3$ & $(\chi_1,\chi_2) \in \{(-0.75,0.75),(0,0.8333)\}$ & ''  \\

\#5 &\texttt{CCATIE}~\cite{Pollney:2007ss}& $q = 1$ & 
$(\chi_1,\chi_2) = (a,a)$, $a \in \{0, 0.2, 0.4, 0.6\}$ & at $r=160 M$  \\

\#6 &''& $q = 1$ & 
$(\chi_1,\chi_2) = (a,-a)$, $a \in \{0, 0.2, 0.4, 0.6\}$ & ''  \\

\#7ab &''& $q = 1$ & 
$(\chi_1,\chi_2) = (\pm 0.6,a)$, $a \in \{\pm 0.3,0,-0.6\}$ &
''  \\ 

\#8 &\texttt{Llama}~\cite{Pollney:2009yz}& $q \in \{1,2\}$ &
$(\chi_1,\chi_2) =(0,0)$ & Null Infinity\footnote{Only the GW
  radiation corresponding to the
 \texttt{Llama} $q=1$ simulation has been extracted at future
 null-infinity using the Cauchy-characteristic method; the $q=2$
 waveform has been extracted at finite radius and extrapolated to
 $r\rightarrow \infty$.} \\ 

\#9 &\texttt{SpEC}~\cite{Scheel-etal-2006:dual-frame}& $q =1$ &
$(\chi_1,\chi_2) =(0,0)$ & at $r \rightarrow \infty$\footnote{Using
  the extrapolation method described
  in~\cite{Scheel:2008rj} with extrapolation order $n=3$.} \\  

            \hline
            \hline
        \end{tabular*}
    \end{center}
\end{table*}

Due to the complexity of this parameter space, most match-filtered
searches for coalescing binaries have so far employed non-spinning
templates, neglecting the effect of the spin by assuming a small,
tolerable loss in
SNR~\cite{Abbott:2007xi,Abbott:2009tt,Abbott:2009qj}. Dedicated
searches for spinning binaries have attempted to model an enlarged
parameter space by using a template family designed to capture the
spin-induced modulations of the gravitational
waveform~\cite{Abbott:2007ai}. In~\cite{Abbott:2007ai} the spin
effects were modeled using unphysical phenomenological parameters;
however, it would be desirable to devise searches for spinning systems
based on strictly physical parameters.
Indeed,~\cite{VanDenBroeck:2009gd} showed that from the point of view
of detection efficiency at a given false-alarm rate, a search based on
non-physical spinning templates is not superior to a non-spinning
search unless specific signal-based vetoes and other tools are
devised. The performance of spinning searches would increase with the
use of templates determined by physical rather than phenomenological
parameters.  That was the motivation for the waveform family presented
in~\cite{Ajith:2009bn}, where as a first step in modeling the full
spinning-binary parameter space, only binaries with non-precessing
spins were considered. Additionally, it is known from PN treatments of
the inspiral~\cite{Vaishnav:2007nm} and from numerical simulations of
the merger~\cite{Reisswig:2009vc} (which though only considers equal
mass systems), that the dominant spin effect on the waveform is from
the \emph{total spin} of the system. Indeed, in~\cite{Ajith:2009bn} it
was found that the effect of the black hole spins can be modeled with
suficient accuracy using only one spin parameter, roughly
corresponding to the total spin of the two black holes. We adopt the
same approach here.  

There are a number of NR simulations of
non-precessing systems for a variety of spin values and mass
ratios. Results with the \texttt{BAM} code are reported
in~\cite{Hannam:2007wf} for the orbital hang-up case and
in~\cite{Hannam:2010aa} for anti-aligned spins. The \texttt{CCATIE}
simulations are presented
in~\cite{Ajith:2009bn,Pollney:2007ss,Rezzolla-etal-2007,Rezzolla:2007rd}; a long
spectral simulation with anti-aligned spins can be found
in~\cite{Chu:2009md}.

\subsubsection{NR waveforms and codes}

The NR waveforms employed in the
construction of the hybrid model used in this paper
are summarized in
Table~\ref{tab:NRWaveforms}.  They have been produced with four
independent NR codes, \texttt{BAM}, \texttt{CCATIE},
\texttt{Llama} and \texttt{SpEC}. The first 3 codes use the
moving-puncture approach~\cite{Baker:2005vv,Campanelli:2005dd} 
to solve the  Einstein equations in a decomposed 3+1 spacetime 
while the last implements the generalized harmonic 
formulation~\cite{Pretorius:2004jg,Boyle:2007ft}. 
\texttt{BAM} and \texttt{CCATIE} use computational domains
based on Cartesian coordinates, while the \texttt{SpEC} code uses a 
sophisticated series of spherical and cylindrical domains; in the wave
zone, the outer computational domains have the same angular
resolution, thus the computational cost only increases linearly with the
radius of the outermost shell. A summary of the properties of
the three codes is given in~\cite{Hannam:2009hh}. 
The \texttt{Llama} code~\cite{Pollney:2009ut,Pollney:2009yz} is based on finite
differencing but the set-up 
of the numerical grid in the outer wave zone is, as in \texttt{SpEC},
also based on spherical coordinates with constant angular
separation. The large wave-zone enables accurate waveform extraction
at large distances, accurate extraction of higher angular modes of the
radiation, and it allows the outer boundary to be far enough away so
that it is causally disconnected from the sphere where the radiation
is extracted.  

The \texttt{BAM} data-set \#1 covers the parameter space of
non-spinning systems for several mass ratios during at least the last
5 orbits before merger (length $\sim 1100-1450\,M$, where $M$ is the
total ADM mass of the spacetime)~
\cite{Hannam:2007ik,Ajith:2007kx,Ajith:2007xh,Hannam:2010aa}.
Data-set \#2 consists of moderately long simulations covering at least
the last 8 orbits before merger (length $\sim 1500-2200\,M$) for
equal-mass systems with equal spins, and are described in depth
in~\cite{Hannam:2007wf,Hannam:2010aa}.  Data-set \#3 consists of
unequal-mass, unequal-spins simulations~\cite{Ajith:2009bn}.  Data-set
\#4 is a simulation with unequal mass and unequal spins employed in
the verification of our fitting mode~\cite{Ajith:2009bn}.  For the
sets \#1--4, initial momenta for quasi-circular orbits were computed
for non-spinning cases according to the procedures described
in~\cite{Husa:2007rh}, leading to low-eccentricity ($e < 0.006$)
inspiral evolutions.  A number of different methods were used for the
spinning cases~\cite{Brugmann:2007zj,Hannam:2007wf,Hannam:2010aa},
depending on which method gave the lowest eccentricity for a given
configuration.  The GW radiation is calculated from the Weyl tensor
component $\Psi_4$ (see e.g. \cite{PenroseRindler}) and extracted at
a sphere with radius $R = 90\,M$. In all cases the uncertainty in the 
phase is less than 0.1\,rad during inspiral (up to $M\omega = 0.1$),
and less than 0.5\,rad during merger and ringdown. The uncertainty 
in the amplitude is less than 0.5\% during inspiral, and less than 5\% during
merger and ringdown.

The \texttt{CCATIE} data-sets \#5, \#6 and \#7ab correspond to the $s$--,
$u$--, $r$-- and $t$--sequences studied
in~\cite{Reisswig:2009vc}. They span the last 
$\sim 4-5$ orbits before merger (length $\sim 500-1000\,M$) and are in
fact not sufficiently long for use in the hybrid construction.
They are still useful to independently verify the reliability of our
phenomenological fit.  Data-set \#5 corresponds to the hang-up
configuration analogous to the \texttt{BAM} set \#1; data-set \#6
consists of configurations with $(\chi_1,\chi_2)=(a,-a)$, i.e. zero
net spin; data-set \#7a was analyzed in~\cite{Pollney:2007ss} in the
context of the study of the recoil velocity (``kick'') of the final
merged black hole.  GW radiation is extracted at $R=160\,M$ via the
Regge-Wheeler-Zerilli formalism for perturbations of a Schwarzschild
black hole
\cite{Regge:1957td,Zerilli:1970se,Abrahams:1988vr,Abrahams:1990jf}. 
  
Data-set \#8 consists of two waveforms for non-spinning black holes
with mass ratios $q=1,2$.  The black holes are evolved with the
\texttt{Llama} code according to the set-up reported
in~\cite{Pollney:2009yz}.  The outer boundary is placed at $3600M$ and
the initial separation is $11M$, corresponding to 8 orbits in the
inspiral phase followed by merger and ringdown.  Wave extraction for
the $q=1$ configuration is done via the Cauchy-characteristic
method~\cite{Reisswig:2009us,Reisswig:2009rx}, taking boundary data
from the numerical spacetime for a subsequent characteristic evolution
of the metric to null-infinity, thereby obtaining waveforms that are
mathematically unambiguous and free of any systematic finite radius
and gauge effects.  The only remaining source of error is due to
numerical discretization.  The equal-mass waveform produced with this
code was reported in \cite{Reisswig:2009rx}, while the $q=2$ waveform
is new.  For these data, the uncertainty in the phase is comparable to
that of the {\tt BAM} waveforms, while the uncertainty in the
amplitude is at least an order of magnitude lower, because of the more
sophisticated wave extraction
procedure~\cite{Pollney:2009yz,Reisswig:2009us,Reisswig:2009rx}.

Data-set \#9 consists of a long non-spinning, equal-mass simulation
that follows 16 orbits of the binary plus merger and ringdown of the
final black hole (length $\sim 4300\,M$). These are publicly
available data~\cite{SpEC:wfs} which were originally computed using
the \texttt{SpEC} code with negligible initial orbital eccentricity
($\sim 5\times10^{-5}$). The GW radiation is extracted via $\Psi_4$
in a similar manner to \#1--4 
and extrapolated to infinity. The phase uncertainty is
less that 0.006\,rad during inspiral, and less than 0.02\,rad during
merger and ringdown; the amplitude uncertainty is less than 0.1\%
during inspiral, and less than 0.3\% during merger and ringdown. A
full description of this simulation is given
in~\cite{Scheel:2008rj}. The long duration of the waveform allows for
its use in the estimation of the errors associated with the length of
the NR data. In particular, since it contains physical information at
lower frequencies, it can be matched to PN results at lower
frequencies, where the PN errors are expected to be smaller (see the
discussion around the right panel of Fig.~\ref{fig:DiffHybs}).

\subsubsection{Going from $\Psi_4$ to $h$}

The gravitational waveforms calculated using NR codes are typically
reported in terms of the Weyl tensor component $\Psi_4$, which is a
complex function that encodes the 
two polarizations of the outgoing transverse radiation.  $\Psi_4$ is
related to the two polarizations of the gravitational wave
perturbation $h_{+,\times}$ (in the transverse-traceless gauge) via
two time derivatives
\begin{equation}
  \label{eq:13}
 \Psi_4 = \frac{d^2}{dt^2} \left[  h_+(t) - ih_\times(t) \right] .
\end{equation}
Going from $\Psi_4$ to $h_{+,\times}$ thus involves two time
integrations and requires us to fix two integration constants
appropriately, corresponding to the freedom to add a linear function
to the strain.  

\begin{figure}
 \centering
\includegraphics[width=\columnwidth]{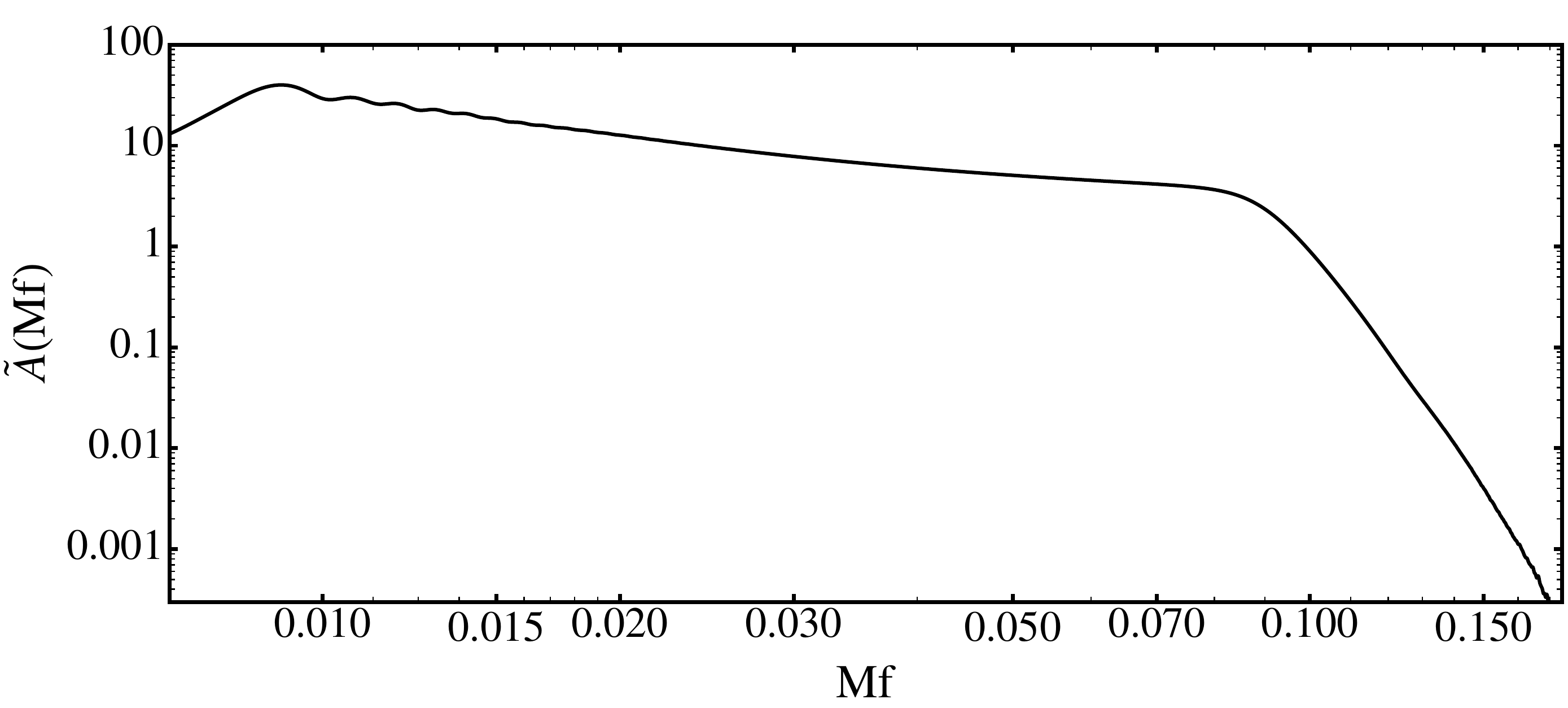}
\includegraphics[width=\columnwidth]{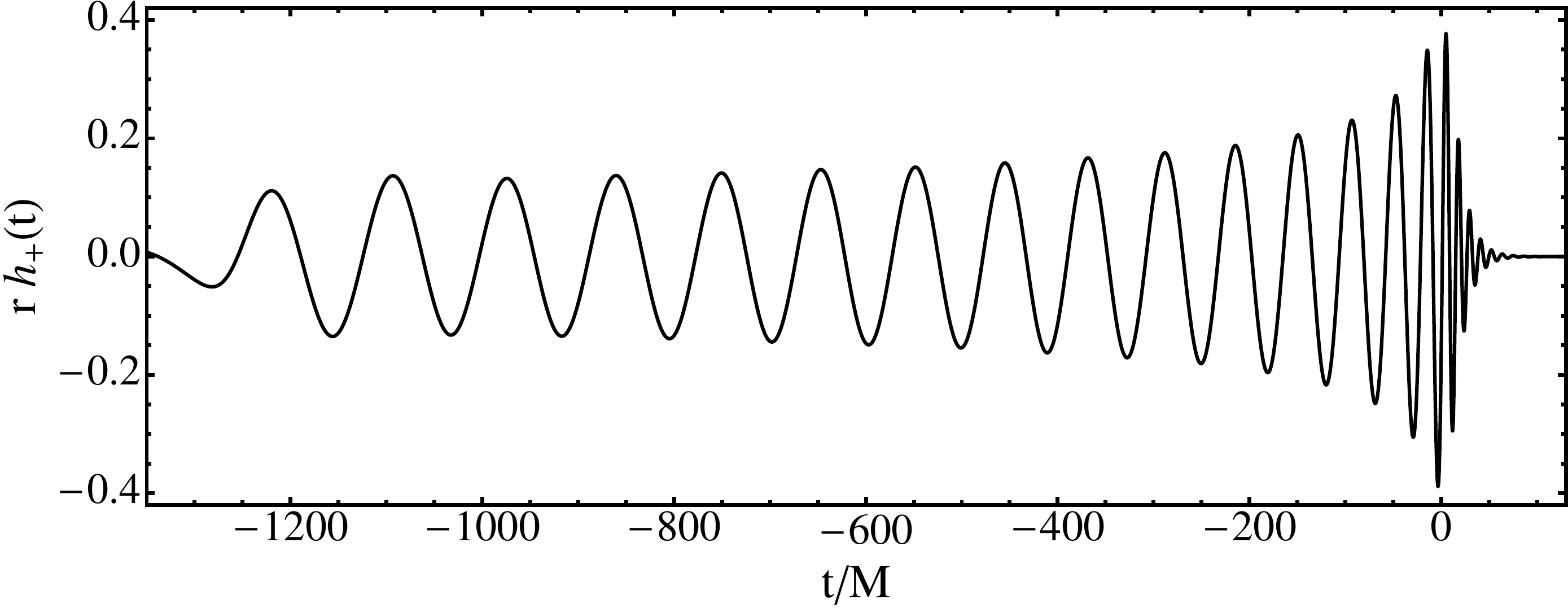}
\caption{ These figures demonstrate the strain waveform obtained by
  the frequency domain division method of calculating $h$ from
  $\Psi_4$. We consider the NR simulation from data-set \#8 of
  Table~\ref{tab:NRWaveforms} with $q=1$, and we start with the
  dominant mode of $\Psi_4$ from this simulation.  The upper
  panel shows $\tilde A = r |\tilde{h}(f)|$ (where $r$ is the extraction radius)
  obtained by the frequency domain
  division, and the lower panel shows $h_+ (t)$.  The window function
  employed by our inverse Fourier transform algorithm is responsible
  for the partial loss of the first cycle of the waveform.
  Nevertheless, a clean $|h(t)|$ during the rest of the inspiral is
  observed.}
\label{fig:IntComp}
\end{figure}

The frequency domain offers a straightforward way of calculating the
strain
\begin{equation}
 h = h_+- ih_\times
\end{equation}
from $\Psi_4$, since integration is replaced by division:
\begin{equation}
\tilde{h}^{\,\rm NR}(f) = - \frac{\tilde{\Psi}_4^{\rm NR}(f)}{4
  \pi^2 f^2} = A^{\rm NR}(f) \, e^{i \Phi^{\rm NR}(f)}\, ,
\label{eq:psi4Toh}
\end{equation}
where $\tilde x(f)$ denotes the Fourier transform of $x(t)$ as defined
in Eq.~\eqref{eq:Fourier_x}. In the limit of large signal
durations, the integration constants only affect the zero-frequency
component of the signal in the frequency domain. For finite duration
signals, the effect of the integration constants will spill over into
higher frequencies like $1/f$.  Since for our purposes, the numerical
simulation provides useful information only starting at a finite
frequency, we conveniently apply a high-pass filter to the data,
thus reducing the effect of the integration constants without using a
fitting procedure.  When performing the division in the frequency
domain, tanh-window functions are employed to pass-filter the data
before computing the Fourier transform.  Fig.~\ref{fig:IntComp}
illustrates the efficacy of our approach for the \texttt{Llama}
equal-mass waveform.  Though we do not discuss it further here, in
general we find that the time- and frequency-domain integration
techniques, both with some fine-tuning, yield comparable results.

\section{Analytical waveforms for spinning binaries using
  the post-Newtonian approach}
\label{sec:postnewton}

\allowdisplaybreaks

Coalescing compact binaries such as BBHs can
be accurately modeled by the PN approximation to general
relativity at least during the major part of the long inspiral phase,
under the assumptions of a weak gravitational field~\cite{Blanchet:LivRev}. 
In order to obtain an analytical description of the early inspiral in
the Fourier domain 
we construct the TaylorF2 phase \cite{Damour:2000zb, Damour:2002kr,
  PhysRevD.72.029901, Arun:2004hn} and the 3PN
amplitude~\cite{Blanchet:2008je,Arun:2008kb} for 
compact binaries with comparable masses and spins (anti-)aligned with the
orbital angular momentum. 

The PN expansion of the binding energy $\mathcal E$ of such systems in
the adiabatic approximation can be taken from the literature, see for
instance  \cite{Blanchet:LivRev, Damour:2001bu, BDEI04,
  Buonanno:2005xu} and references therein. For the results shown here
we include leading order and next-to-leading order spin-orbit effects
\cite{Kidder:1995zr, Apostolatos:1994mx, Blanchet:2006gy} as well as
spin-spin effects that appear at relative 2PN order
\cite{Kidder:1995zr, Damour:2001tu, Poisson:1997ha}; note that the
square terms in the individual spins are valid only for black holes as
discussed in \cite{Damour:2001tu, Poisson:1997ha,
  Buonanno:2005xu}. The notation used in this section adopts unit
total mass $M =1$ and $G = c =1$. Each black hole is characterized by
its mass $m_i$ and the magnitude of its spin 
\begin{equation}
S_i = \vert \chi_i \vert \, m_i^2,\qquad i = 1,2.
\label{eq:spin}
\end{equation}
The spin vectors are (anti-)aligned with the
orbital angular momentum $\bm{L}$, where the sign of $\bm{L} \cdot
\bm{S_i}$ defines the sign of $\chi_i$. 
With the aim of matching to available NR data, we use the PN
spin definition that yields constant spin magnitudes \cite{Faye:2006gx,
  Blanchet:2006gy}. 
The quantity 
\begin{equation}
 \eta = \frac{m_1 \: m_2}{M^2} \label{eq:symmmassratio}
\end{equation}
is the symmetric mass ratio. The PN expansion is written in the
dimensionless variable $x$, related to the orbital angular frequency
$\omega$ of the binary via $x = \omega^{2/3}$. To summarize the structure
of this derivation, we start by giving the energy for the considered
scenario as
\begin{equation}
  \mathcal E = -\frac{x \eta}{2} \sum_{k=0}^6 e_k \:
  x^{k/2},  \label{eq:PNenergy} 
\end{equation} 
where the coefficients $e_k$ are listed in Eq.~\eqref{eq:energyCoeff}. 

The other ingredient needed to describe an inspiraling BBH as a
sequence of quasi-circular orbits is the flux $\mathcal{F}$, which we
take at 3.5PN order including the same spin effects as for the
energy. We additionally take into account the 2.5PN correction of the
flux due to the energy flow into the BHs, calculated in
\cite{Alvi:2001mx}. The final result is 
\begin{equation}
 \mathcal F = \frac{32}{5}\eta^2 x^5 \sum_{k=0}^7 f_k \: x^{k/2}~, \label{eq:flux}
\end{equation}
where the coefficients $f_k$ are given in Eq.~\eqref{eq:fluxCoeffs}.

The energy loss of the system due to gravitational radiation is
expressed as \mbox{$d \mathcal E (t)/dt = - \mathcal F(t)$}, which
translates to 
an evolution equation for the orbital frequency, or equivalently 
\begin{equation}
\frac{dx}{dt} = - \frac{\mathcal F (x)}{d \mathcal E(x)/dx}
~. \label{eq:xdot_FE} 
\end{equation}
Starting from (\ref{eq:xdot_FE}), different waveform models can be
constructed, for overviews see
\cite{Boyle:2007ft,Buonanno:2009zt}. For the purpose of the results
shown here,
we shall give the relevant expressions in the frequency domain later
and explicitly construct only the TaylorT4 approximant, which is
obtained by expanding the right-hand side of Eq.~\eqref{eq:xdot_FE} to
3.5PN order
\begin{equation}
 \frac{dx}{dt} = \frac{64}{5}\eta x^5 \sum_{k=0}^7 a_k \: x^{k/2}~, \label{eq:T4}
\end{equation}
with $a_k$ given in~(\ref{eq:T4Coeffs}).

Note that the formal re-expansion of the denominator and the
multiplication with the numerator in Eq.~\eqref{eq:xdot_FE} also yields
contributions to higher orders than those in Eq.~\eqref{eq:T4}. However,
since 4PN and higher terms in flux and energy are not fully
determined, the expressions one can compute for $a_k$ with $k>7$ are
incomplete. The same applies to contributions of the spins at
relative PN orders higher than 2.5PN. When we later use the TaylorT4
expression (\ref{eq:T4}) in this paper, we only expand it to 3.5PN
order but keep all the spin terms that appear, i.e.
{\em incomplete} contributions in $a_6$ and $a_7$ are not
neglected. Only if higher order spin corrections at 3 and 3.5PN order become available in 
energy and flux, the corresponding spin terms in the TaylorT4 (and TaylorF2)
description can be completed.

In order to construct an analytical formula of the wave signal in the Fourier
domain, the stationary phase approximation is commonly used to obtain
the TaylorF2 expression for the phase \cite{Damour:2000zb,
  Damour:2002kr, PhysRevD.72.029901, Arun:2004hn}. Below, we briefly
recapitulate the steps towards the derivation of this approximant and
provide the final result.  

Expanding the inverse of relation~(\ref{eq:xdot_FE}), 
$dt/dx = - (d\mathcal E / dx)/\mathcal F$, allows for the analytical
integration of 
$t(x)$. The orbital phase $\phi$ can be integrated via 
\begin{equation}
 \frac{d \phi}{dt} = \omega = x^{3/2} \quad \Rightarrow \quad \frac{d
   \phi}{dx} = - x^{3/2}\: \frac{d \mathcal E(x) /dx }{\mathcal F(x)} 
\end{equation}
to obtain $\phi(x)$. This is the definition of the TaylorT2
approximant. The $(\ell,m)$ modes of the decomposition of the gravitational
radiation in spherical harmonics can be approximated in the time domain
by \cite{Blanchet:2008je} 
\begin{equation}
 h^{\ell m} (t) = A^{\ell m}(t) \: e^{- i m \phi(t)} ~,
\end{equation}
and the transformation to the frequency domain is carried out in the
framework of the stationary phase approximation
\begin{align}
 \tilde h^{\ell m}(f) &= \int_{-\infty}^{\infty} h^{\ell m} (t) \: e^{2 \pi i ft} dt \\
 & \approx A^{\ell m} (t_f) \sqrt{\frac{2 \pi}{m \ddot \phi (t_f)} } \:
 e^{i \psi^{\ell m}(f)} , \label{eq:h_SPA} 
\end{align}
where $t_f$ is defined as the moment of time when the instantaneous
frequency coincides with the Fourier variable, i.e., $m \omega(t_f) =
2 \pi f$. The phase in the frequency domain is given by 
\begin{equation}
 \psi^{\ell m} (f) = 2 \pi f \, t_f - m \phi (t_f) - \frac{\pi}{4}
 ~. \label{eq:psi_lm} 
\end{equation}
Given $t(x)$ and $\phi(x)$ one can immediately change to the Fourier
variable by
\begin{equation}
 x(t_f) = \left[\,\omega(t_f)\,\right]^{2/3} = \left( \frac{2 \pi
     f}{m} \right)^{2/3}. \label{eq:x_tf} 
\end{equation}
Starting from the energy~(\ref{eq:PNenergy}) and flux~(\ref{eq:flux})
consequently leads to  
\begin{align}
\begin{split}
\psi^{22}(f) &= 2 \pi f t_0 - \phi_0 - \frac{\pi}{4} \\&  + \frac{3}{128 \eta } (\pi f)^{-5/3} \sum_{k=0}^7 \alpha_k (\pi f)^{k/3},
\end{split}
\label{eq:F2phase} 
\end{align}
with the corresponding coefficients $\alpha_k$ of~(\ref{eq:F2Coeffs}).
From (\ref{eq:psi_lm}) one realizes that, in fact, Eq.~\eqref{eq:F2phase}
is valid for all spherical harmonics with $m = 2$. The quantities $t_0$
and $\phi_0$ are arbitrary and arise as integration constants when
calculating $t(x)$ and $\phi(x)$. 
When implementing this Fourier domain phase we also take
into account the spin terms that appear after re-expanding at 3PN and
3.5PN order, although they are not complete.

The time-domain amplitude of the gravitational wave was recently
calculated at 3PN order by Blanchet et al.~\cite{Blanchet:2008je}. We
use the expression given by them for the $\ell=2$, $m=2$ mode in
combination with the spin corrections provided in
\cite{Berti:2007nw,Arun:2008kb}. In our notation, the time-domain
amplitude reads 
\begin{equation}
    A^{22} (x) = \frac{8 \eta \, x }{D_L} \sqrt{\frac{\pi}{5}}
    \sum_{k=0}^6 \mathcal A_k \, x^{k/2} , \label{eq:Amp_time} 
\end{equation}
where $D_L$ is the luminosity distance between source and observer and
the coefficients $\mathcal A_k $ are given in~(\ref{eq:ampCoeffs}). 

\begin{figure}[tbp]
 \centering
\includegraphics[width=0.95\columnwidth]{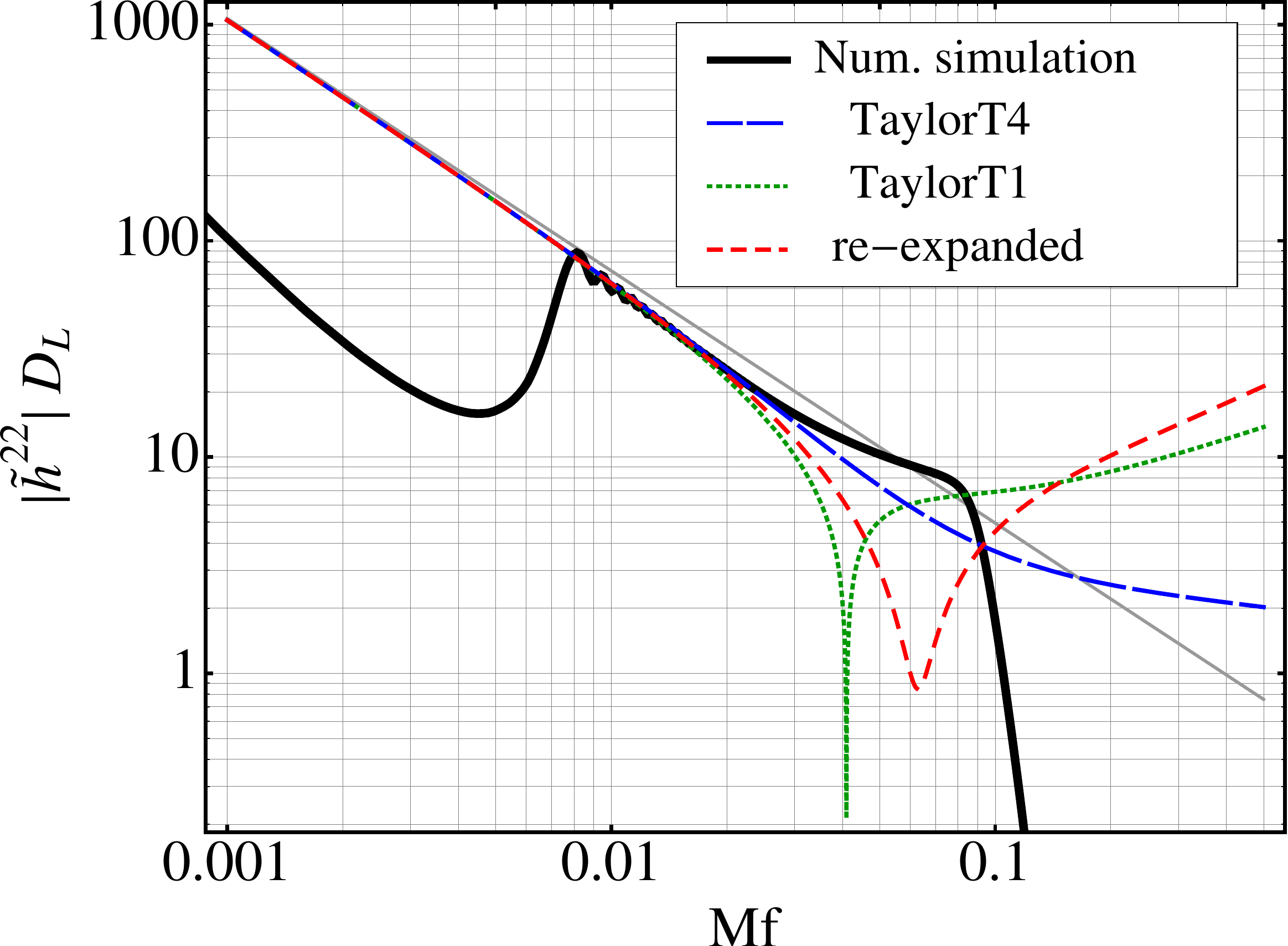}
\caption{Different variants of constructing the PN Fourier amplitude
  in the stationary phase approximation for the equal-mass case. The
  labels explain how $(\pi/\ddot \phi)^{1/2}$ is treated in
  (\ref{eq:h_SPA}). The thick curve shows data obtained by a numerical
  simulation in full general relativity which begins at $Mf \approx
  0.008$. The straight gray line illustrates the restricted PN
  amplitude, $\vert \tilde h^{22} \vert 
  D_L = \pi \sqrt{2 \eta /3 } (\pi f)^{-7/6}$. } 
\label{fig:PNamps}
\end{figure}

From (\ref{eq:h_SPA}) we see that, in order to construct the Fourier domain
amplitude, an explicit expression for $\ddot \phi = d^2 \phi/dt^2 = \dot
\omega$ is needed. In \cite{Arun:2008kb} this is done by re-expanding
$\sqrt{1/\dot \omega}$ using the same ingredients as those underlying
the TaylorT$n$ approximants. We may, however, look at $\dot \omega =
(3/2) \sqrt{x} \, \dot x$ and choose one ``preferred'' prescription for
$\dot x$ without re-expanding the quotient. Aiming at matching
PN results to NR waveforms, we compare
different possibilities of replacing $\dot x$ (namely by its TaylorT1
and TaylorT4 description) and the re-expansion of the form 
\begin{equation}
 \sqrt{\frac{\pi}{\dot \omega}} \approx \sqrt{\frac{5 \pi}{96 \eta} }
 x^{-11/4} \sum_{k=0}^7 \mathcal S_k \: x^{k/2} 
\end{equation}
(see \cite{Arun:2008kb}) with data of numerical simulations in
full general relativity. The result in the equal-mass
case  can be seen in Fig.~\ref{fig:PNamps}. Note that the transfer to the 
Fourier domain is completed by using (\ref{eq:Amp_time}) in
(\ref{eq:h_SPA}) in combination with (\ref{eq:x_tf}).  

All variants of the 3PN Fourier amplitude agree reasonably well with the
numerical relativity data roughly up to the frequency of the last
stable circular orbit in the Schwarzschild limit, $Mf = \pi^{-1} \, 6^{-3/2}
\approx 0.022$. Due to a comparable behavior even beyond this point we
choose to construct the Fourier amplitude of our post-Newtonian model
by using the TaylorT4 $\dot x$~(\ref{eq:T4}). The same choice was
employed e.g. in \cite{Read:2009yp}.

\section{Matching post-Newtonian and numerical relativity waveforms}
\label{sec:pn-vs-nr}

\subsection{Basic notions}

The basic criteria for evaluating the goodness of fit for the hybrid
waveform require a notion of distance between two GW signals
$h(t)$ and $h^\prime(t)$.  The simplest notion is the distance in the
least-squares sense over an interval $t_1\leq t \leq
t_2$~\cite{Ajith:2007qp,Ajith:2007kx,Ajith:2007xh}
\begin{equation}
  \label{eq:6}
  \delta_{t_1,t_2}(h,h^\prime) = \int_{t_1}^{t_2} \left|h(t) -
    h^\prime(t)\right|^2\,dt\,. 
\end{equation}
This can be used for the numerical relativity $h_{\rm NR}(t)$
and the post-Newtonian waveform $h_{\rm PN}(t)$, with the interval
$[t_1,t_2]$ being chosen so that both waveforms are reasonably good
approximations (in a sense to be quantified later).  Thus, the PN
waveform is taken up to $t_2$ and the NR waveform is taken to start at
$t_1$, and they overlap within the interval $[t_1,t_2]$. 

Let us consider the frequency domain equivalent.  Our convention
for the Fourier transform of a signal $x(t)$ is
\begin{equation}
  \label{eq:Fourier_x}
  \tilde{x}(f) = \int_{-\infty}^\infty x(t) \, e^{ 2\pi ift}dt\,.
\end{equation}
One needs to be careful in converting the time interval $[t_1,t_2]$ to a
frequency interval $[f_1,f_2]$.  In principle, the Fourier transform
is ``global'' in time; signals that have compact support in
time cannot have compact support in frequency, and vice versa.
However, for the binary black hole waveforms (prior to the ring-down stage) that we are considering,
the frequency always increases in time, so that we can sensibly
associate a frequency interval $[f_1,f_2]$ with a given time interval
$[t_1,t_2]$. For these waveforms, we can consider the above distance
definition in the frequency domain:
\begin{equation}
  \label{eq:7}
  \delta_{f_1,f_2}(h,h^\prime) = \int_{f_1}^{f_2}\left| \tilde{h}(f) -
  \tilde{h}^\prime(f)\right|^2df\,.
\end{equation}
We shall use such a norm (applied to the phase) for constructing the
hybrid waveform.  

When evaluating the goodness of a hybrid waveform for a particular
detector, we need to consider detector-specific inner products, which
are convenient to describe in the frequency domain.  Let $S_n(f)$ be
the single-sided power spectral density of the noise in a GW detector
defined as
\begin{equation}
  \label{eq:3}
  \mathbf{E}\left[\tilde{n}(f)\tilde{n}^\ast(f^\prime)\right] =
  \frac{1}{2}S_n(f) \, \delta(f-f^\prime)\,. 
\end{equation}
Here $n(t)$ is the detector noise time series with $\tilde{n}(f)$ its
Fourier transform, $^\ast$ denotes complex conjugation and
$\mathbf{E}$ refers to the expectation value over an ensemble of
independent realizations of the noise, which is assumed to be a
zero-mean, stationary, stochastic process.  This equation implies that
data at different frequencies are independent, and is one of the
reasons why working in the frequency domain is so useful in data
analysis.  The time domain description of the noise is more
complicated; $n(t)$ and $n(t+\tau)$ are in general not independent;
$\mathbf{E}[n(t)n(t+\tau)]$ is generally non-zero.  For stationary
noise this is a function $C(\tau)$ only of $\tau$, and is related to
$S_n(f)$ via a Fourier transform (see e.g. \cite{Papoulis}).

Given $S_n(f)$, we use the following definition of an inner product
between two signals $x(t)$ and $y(t)$
\begin{equation}
   \label{eq:1}
   (x|y) \equiv 4 \mathrm{Re}\int_0^\infty 
   \frac{\tilde{x}(f)\tilde{y}^\ast(f)}{S_n(f)}df ,
\end{equation}
where $\tilde{x}(f),\tilde{y}(f)$ are the Fourier transforms of $x(t),
y(t)$ respectively. This inner product is appropriate for Gaussian
noise and forms the basis for matched filtering (see
e.g. \cite{Helstrom}).  It can be used to define a suitable notion of
distance between two signals $h(t)$ and $h^\prime(t)$ as $(\delta
h|\delta h)^{1/2}$, where $\delta h(t) = h^\prime(t) - h(t)$.

The distinguishability between $h(t)$ and $h^\prime(t)$ in the
presence of noise can be understood with the following construction.
Following \cite{Lindblom:2008cm}, we define a 1-parameter family of
waveforms which interpolates linearly between $h(t)$ and $h^\prime(t)$
as
\begin{equation}
  \label{eq:4}
  h^{\prime\prime}(t;\lambda) = h(t) + \lambda \, \delta h(t) \,.
\end{equation}
We obviously have $h^{\prime\prime}(t;0) = h(t)$ and
$h^{\prime\prime}(t;1) = h^{\prime}(t)$.  The question of
distinguishability between $h(t)$ and $h^\prime(t)$ now becomes one of
estimating the value of $\lambda$ [for the extended signal model
$h^{\prime\prime}(t;\lambda)$] in the presence of noise.  If we use an
unbiased estimator for $\lambda$, the variance $\sigma_\lambda^2$ of
the estimator is bounded from below by the Cramer-Rao bound (see
e.g. \cite{Kendalls})
\begin{equation}
  \label{eq:5}
  \sigma_\lambda^2 \geq (\delta h|\delta h)^{-1}\,.
\end{equation}
This can be a useful bound for large SNRs, which is in fact what we are
interested in here; it is easier to distinguish between two loud
waveforms and demands on the waveform model are correspondingly more
stringent.  Thus, a useful condition for being able to distinguish
between the two waveforms is $\sigma_\lambda < 1$.  If $h(t)$ is
the true waveform and $h^{\prime}(t)$ our approximation to it, then we
say that $h^{\prime}(t)$ is a sufficiently accurate approximation if
$(\delta h|\delta h) \leq 1$.  Assuming that $(h|h) \approx (h^\prime|h^\prime)$,
 it is clear that $(\delta h |\delta h)
\propto \rho^2$ where $\rho = (h|h)^{1/2}$ is the optimal SNR.  Hence,
as we just remarked, the two signals are easier to distinguish when
the detector is more sensitive, or when the signal amplitude is
larger.  It will be convenient to normalize the norm of $\delta h$ and
write this distinguishability criterion as
\begin{equation}
  \label{eq:11}
  \frac{1}{\rho^2}(\delta h|\delta h) \geq \frac{1}{\rho^2}\,.
\end{equation}
Thus, for a given detector, we choose a reasonable guess $\rho_0$ for
the largest expected SNR and we compute the
normalized distance between the two waveforms $(\delta h|\delta
h)/\rho^2$.  If this exceeds $1/\rho_0^2$, then we consider that the
detector is able to distinguish between the two waveforms.

If we are interested in the less stringent requirement of detection
rather than in strict distinguishability, then a sufficient condition
is \cite{Lindblom:2008cm}
\begin{equation}
  \label{eq:14}
  \frac{1}{\rho^2}(\delta h|\delta h) < 2\epsilon,
\end{equation}
where $\epsilon$ is the maximum tolerated fractional loss in SNR.
More explicitly: if $h(t)$ is the exact waveform and $h^\prime(t)$ an
approximation thereof, then the approximation is potentially useful
for detection purposes if \eqref{eq:14} is satisfied for an
appropriate choice of $\epsilon$.  If we are willing to accept e.g. a
10\% loss in detection rate, then a suitable choice is $\epsilon
\approx 0.10/3 \approx 0.03$ (corresponding to sources uniformly
distributed in space).  This value of $\epsilon$ does not take into
account the additional loss in SNR due to discrete template banks used
in realistic searches. In practice one might need to choose $\epsilon$
an order of magnitude smaller than this so that the total fractional
loss in SNR remains acceptable.  Since a more precise value is
pipeline dependent, we shall ignore this caveat and use $\epsilon =
0.03$ as a convenient reference; the reader can easily scale the
results of this paper appropriately for different choices.

A useful way to describe the efficacy of approximate waveform models
is through the concepts of effectualness and faithfulness introduced
in \cite{DIS98}.  Let $h_\lambda(t)$ be the exact waveform
with parameters $\lambda$ and the approximate waveform model be
$h_\lambda^{\rm app}(t)$. The ambiguity function is defined as the
normalized inner product maximized over extrinsic parameters
\begin{equation}
  \label{eq:12}
  \mathcal{A}(\lambda,\lambda^\prime) = \max_{t_0,\phi_0}
  \frac{(h_{\lambda}|h^{\rm
        app}_{\lambda^\prime})}{\sqrt{(h_\lambda|h_\lambda)(h^{\rm
        app}_{\lambda^\prime}|h^{\rm app}_{\lambda^\prime})}} ,
\end{equation}
where $t_0$ is the time offset between the two waveforms, and $\phi_0$
is the initial phase. Performing a further maximization over the
parameters $\lambda^\prime$ of the model waveforms, we define
$\hat{\mathcal{A}}(\lambda) =
\max_{\lambda^\prime}\mathcal{A}(\lambda,\lambda^\prime)$.  If
$\hat{\mathcal{A}}(\lambda)$ exceeds a chosen threshold, e.g. 0.97,
then the waveform model $h^{\rm app}$ is said to be
{\em effectual}. Effectual models are sufficient for detection. In
order to be 
able to estimate parameters we also need the model to be {\em faithful}.
This means that the value of $\lambda^\prime$ which maximizes
$\mathcal{A}(\lambda,\lambda^\prime)$ should not be biased too far
away from $\lambda$.

\subsection{Issues in matching PN with NR}

It is useful at this stage to discuss some of the issues that arise in
combining PN and NR results.  The discussion here will be short and
incomplete, and the topic merits an in-depth investigation that is
beyond the scope of the present work. Our immediate aim is simply to
spell out some of the reasons why black hole parameters in the PN and
NR frameworks may not necessarily refer to the same physical
quantities.  One should therefore not be surprised that when combining
NR and PN waveforms, it might become necessary to vary the intrinsic
black hole parameters as well.  This is not to say that either PN or
NR use incorrect definitions for black hole parameters, both
frameworks are in fact consistent within their domains of
applicability.  The point rather is that the two formalisms are quite
different when viewed as approximation schemes to general relativity,
and these differences might need to be taken into account depending on
the accuracy requirements for the matching.

Since PN and NR are both used to address the BBH problem, one could
imagine starting with the two black holes very far apart, evolve them
using appropriate PN equations of motion and compute the resulting
waveforms.  As one gets close to the merger, terminate the PN
evolution and use this end-point to construct initial data for the
full NR simulation which then evolves the black holes through the
merger and ringdown.  However, the formalisms and methods employed in
the two cases are radically different and there are potential
difficulties in carrying out this procedure.  

The PN formalism is based on a perturbative expansion in powers of the small
parameter $\epsilon = v/c$, where $v$ is the orbital velocity and $c$
is the speed of light.  In the usual formulations, PN theory uses a
point-particle description of the black holes, and their parameters
can be viewed as effective parameters which couple in the appropriate
manner with the external background gravitational field (see
e.g. \cite{Goldberger:2004jt,Porto:2008tb}).  The goal of PN theory is
to find a 1-parameter sequence of solutions to the field equations
$g_{\mu\nu}^\epsilon$ to any specified order in $\epsilon$. It has
recently been shown rigorously \cite{Oliynyk:2009qa} that, in the
cosmological setting with gravitating perfect fluids, the 1-parameter
family of solutions exists and admits an expansion in $\epsilon^n$ to
any order.  While similar results in the asymptotically-flat case are
not yet available, it is certainly reassuring to know that PN works
well in this non-trivial setting (in fact, it can be persuasively 
argued that the cosmological setting is more relevant to GW
observations than strict asymptotic flatness).  The errors in PN
waveforms are then due to the systematic differences between the true
waveform and the asymptotic series expansion in $\epsilon^n$ truncated
at a finite order, and this error depends on which particular PN
expansion one chooses to use.

In contrast, numerical relativity is based on the 3+1 formulation of
general relativity as an initial value problem, and one solves the
resulting partial differential equations numerically.  The GW signal
is typically measured at a large, but finite, 
coordinate distance from the source, and encoded in $\Psi_4$, the 
frame-dependent outgoing transverse component of the
Weyl tensor component. The data from multiple coordinate
radii are extrapolated to asymptotic distances, or evaluated at
null infinity in the case that characteristic extraction
is used~\cite{Reisswig:2009us,Reisswig:2009rx}.
For a given
physical configuration (choice of masses, spins, separation etc.), one
specifies the initial data consisting of the spatial metric and
extrinsic curvature of the initial spatial slice.  The physical
initial data parameters
should be chosen to be as compatible as possible with the spacetime
computed in the PN formalism, and significant progress has been made
in this regard~\cite{Husa:2007hp,Walther:2009ng}.  

The black holes
here are not point particles but rather black hole horizons. The
parameters of the black hole are often computed as integrals over the
apparent horizon, and in most cases the parameters used in
constructing the initial data are also useful approximations to the
true ones.  There are, however, possible systematic errors.  For
example, if we are using the quasi-local horizon definitions, an
important requirement is that the horizon should locally be
approximately axisymmetric.  The methods for finding the approximate
symmetry vectors have become increasingly accurate and reliable
\cite{Dreyer:2002mx,Jasiulek:2009zf,Beetle:2008yt,Harte:2008xt,Lovelace:2008tw}.
However, it should be kept in mind that the assumption of approximate
axisymmetry is expected to become increasingly worse closer to the
merger.  Furthermore, the very use of apparent horizons is gauge
dependent; using a different time coordinate will lead to a different
set of apparent horizons and possibly also different values of the
parameters.  In the inspiral phase when the horizons are sufficiently
isolated this gauge issue is not expected to be a problem, but as we
get closer to the merger (this has not yet been quantified), the
variation in the parameters due to gauge choices could become
significant~\cite{Nielsen:2010}.

Let us elaborate a little more on the spin.  Most post-Newtonian
treatments are based on the equations of motion derived in
\cite{Corinaldesi:1951pb,Papapetrou:1951pa}.  The starting point is
the spin tensor $S^{\mu\nu}$ constructed from moments of the stress
energy tensor $T^{\mu\nu}$. Since $S^{\mu\nu}$ has potentially 6
non-zero independent components, the system for the 4 equations of
motion $\nabla_{\mu}T^{\mu\nu} = 0$ is over-determined.  One thus
imposes the additional spin supplementary conditions such as
$S^{\mu\nu}p_{\nu} = 0$ or $S^{\mu\nu}u_{\nu} = 0$ with $p_{\mu}$
being the 4-momentum and $u_{\nu}$ the 4-velocity.  These different
conditions lead to physically different equations of motion and
trajectories \cite{2007MNRAS.382.1922K}.  On the other hand, for black
holes in NR, a common method for evaluating spin employs the formalism
of quasi-local horizons \cite{Ashtekar:2004cn}.  The final result for
the magnitude of the horizon angular momentum is an integral over the
apparent horizon $S$:
\begin{equation}
  \label{eq:8}
  J = -\frac{1}{8\pi}\oint_S K_{\mu\nu}\phi^\mu dS^\nu\,,
\end{equation}
where $K_{\mu\nu}$ is the extrinsic curvature of the Cauchy slice,
$\phi^\mu$ is a suitable approximate axial symmetry vector on $S$
\cite{Dreyer:2002mx,Jasiulek:2009zf,Beetle:2008yt,Harte:2008xt,Lovelace:2008tw},
and $dS^b$ is the area element on the apparent horizon.  The direction
of the spin is harder to find, but some approximate methods are available
\cite{Campanelli:2006fy,Jasiulek:2009zf}.  There is yet no detailed
study of possible analogs of the spin supplementary conditions in this
formalism, or on the equations of motion for horizons with a given set
of multipole moments.  For a horizon with area $A$ and spin
magnitude $J$, the mass is given by the Christodoulou formula
\begin{equation}
  \label{eq:10}
  m = \sqrt{\frac{A}{16\pi} + \frac{4\pi J^2}{A}} \, .
\end{equation}
Hence, uncertainties in spin can also lead to uncertainties in the
mass.    

As long as we are dealing with just the numerical or PN waveforms by
themselves, these small effects in the definitions of mass and spin
are not important for most applications. In fact, we can treat them as
just convenient parameterizations of the waveform without worrying
about their detailed physical interpretation.  However, when we wish
to compare the results from frameworks as different as PN and NR this
may no longer work.  Depending on the details of the matching
procedure, systematic differences between the various definitions
might need to be taken into account, or at the very least they should
be quantified.  If a particular case requires matching a very long PN
portion (depending on the total mass and the lower-frequency cutoff of
a particular detector), then even a small change in the PN parameters
at the matching frequency can translate into a large phase difference
at lower frequencies.  One valid approach is to not assume \emph{a
  priori} that the PN and NR parameters are equal to each other but
rather, for a given numerical waveform, we search over PN waveforms in
a particular PN approximant and find the best fit values.  Finally,
given that the NR and PN parts have different values of physical
parameters, it is a matter of convention what values are to be
assigned to the hybrid.  The values of the parameters in the early
inspiral are a convenient and astrophysically relevant choice.

\subsection{An illustration for non-spinning systems}
\label{sec:errors} 

Let us now move to a concrete case of constructing hybrid waveforms,
considering the non-spinning \texttt{Llama} waveforms, i.e. data set
\#7 in Table~\ref{tab:NRWaveforms}.  Recall that this data set
consists of two waveforms with non-spinning black holes with mass
ratios 1:1 (used in left and central panels of Fig.~\ref{fig:DiffHybs})
and 1:2 (Figs.~\ref{fig:phaseError}, \ref{fig:etaMerr},
\ref{fig:etaVal} and right panel of \ref{fig:DiffHybs}).  Since these waveforms are
calculated using the \texttt{Llama} code with extraction at future
null-infinity with the Cauchy-characteristic method for the equal mass
case, or well into the wave-zone for the 1:2 case, we are confident
that systematic effects of waveform extraction are
small.  Even for these waveforms, based on the discussion above, in
principle we should not rule out a small mismatch in the values of the
spin (and perhaps also eccentricity) between the NR and PN waveforms.
For simplicity, let us consider only the possibility that the
symmetric mass ratio $\eta$ could be different, and restrict ourselves
to non-spinning black holes and zero eccentricity.  We would like to
match the \texttt{Llama} waveforms with the frequency domain PN
waveforms discussed in Sec.~\ref{sec:postnewton} with the values of
the spins set to zero. The total mass $M$ sets the scale for the time
(and frequency); in addition we have the extrinsic parameters for the
time offset and initial phase $t_0$ and $\phi_0$.  Furthermore, we
only consider the $\ell=m=2$ mode, so that the PN waveform is of the
form $\tilde{h}^{\rm PN}(Mf;\phi_0,t_0,\eta_{\rm PN})$ in the
frequency domain.

\subsubsection{Fitting errors}

For a given NR waveform $h_{\rm
  NR}(t)$ we consider a time window $(t_0,t_0 + \Delta t)$ or,
alternatively, in the frequency domain the matching region consists of
a lower starting frequency $f_{L}$ and a width $\Delta f$.  We match the two
waveforms in a least-squares sense by minimizing the phase difference in Fourier space
\begin{align}
  \label{eq:9}
      \delta &= \min_{t_0,\phi_0,\eta_{\rm PN}}\int_{f_L}^{f_L+\Delta
      f}
    \left|\delta\phi(f;\eta_{\rm NR},\eta_{\rm PN},t_0,\phi_0)\right|^2 Mdf\,, \nonumber\\
    \delta\phi(f) &\equiv \phi_{\rm NR}(f;\eta_{\rm NR}) -
    \phi_{\rm PN}(f;t_0,\phi_0,\eta_{\rm PN})\,.
\end{align}
We optimize $\delta$ over all allowed time and phase shifts, i.e.
$(t_0,\phi_0)$, and the PN intrinsic parameters $\lambda_{\rm PN}$.
Given the previous discussion on the possible differences between the
intrinsic parameters $\lambda$ in the PN and NR frameworks, here
we have distinguished between the intrinsic parameter
$\eta$~(\ref{eq:symmmassratio}) appearing in $h_{\rm PN}$ and $h_{\rm
  NR}$. Note that we are not only neglecting spins and eccentricity
but also assume $M_{\rm PN} = M_{\rm NR} = M$.  Future analyses should
successively drop these simplifications.

\begin{figure}
\vspace{0.4cm}
  \includegraphics[width=0.85\columnwidth]{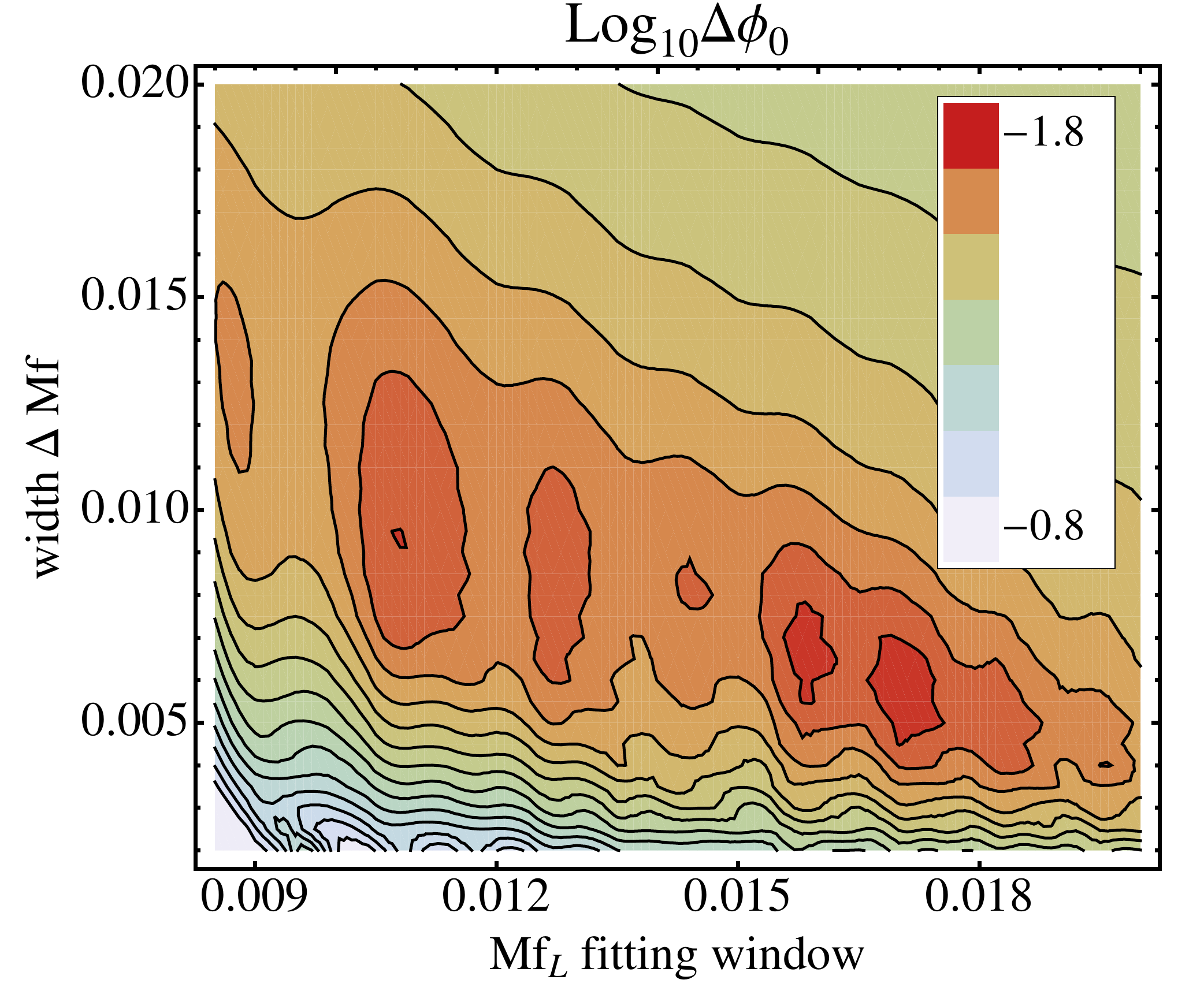}
  \caption{A contour plot for the fitting error $\Delta\phi_0$ in the
    $(f_L,\Delta f)$ plane. Here $\eta$ is kept fixed to the NR value
    and we optimize over $\phi_0$ and $t_0$.}
  \label{fig:phaseError}
\end{figure}

\begin{figure*}
  \includegraphics[width=0.3\textwidth]{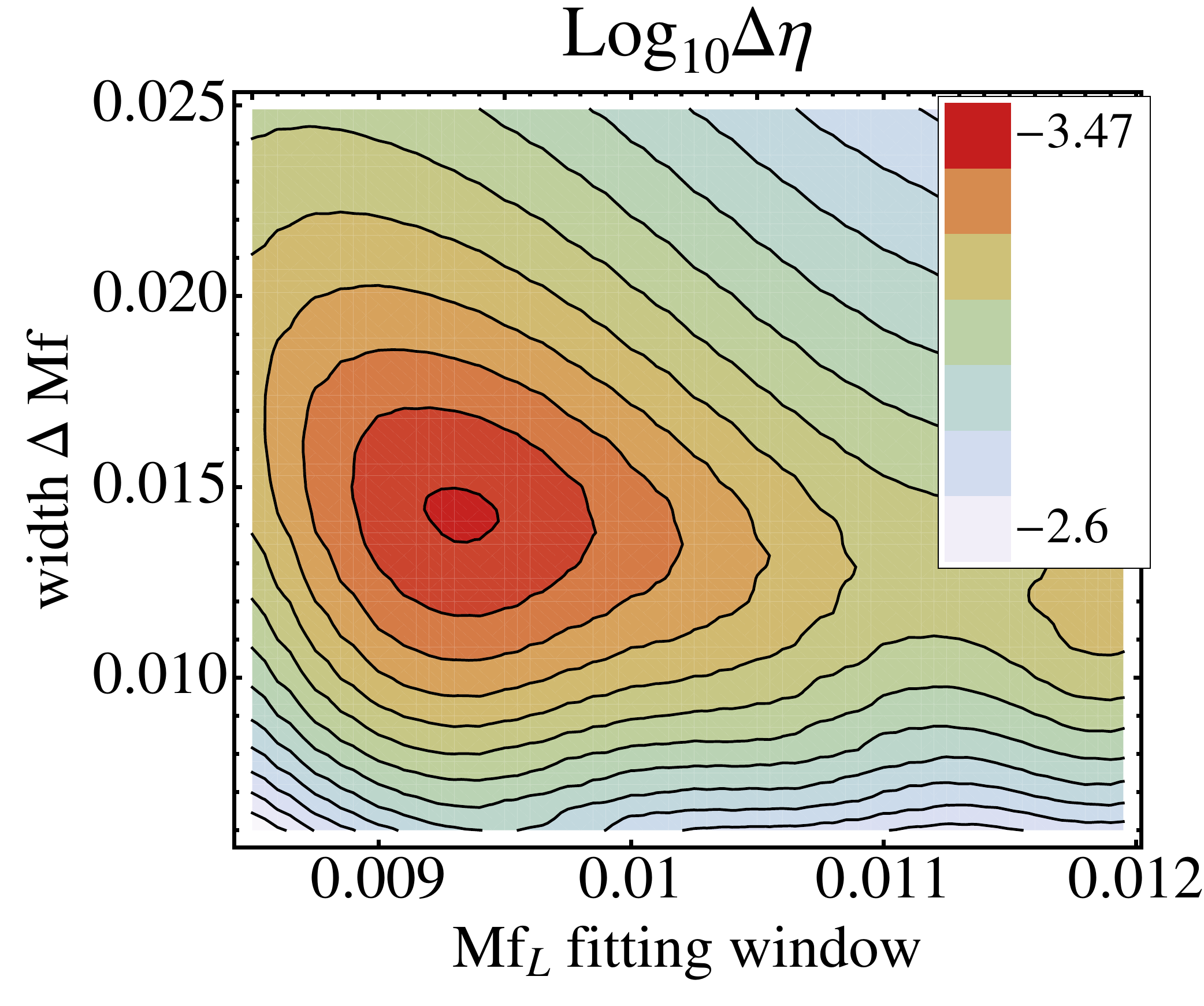} ~ \includegraphics[width=0.3\textwidth]{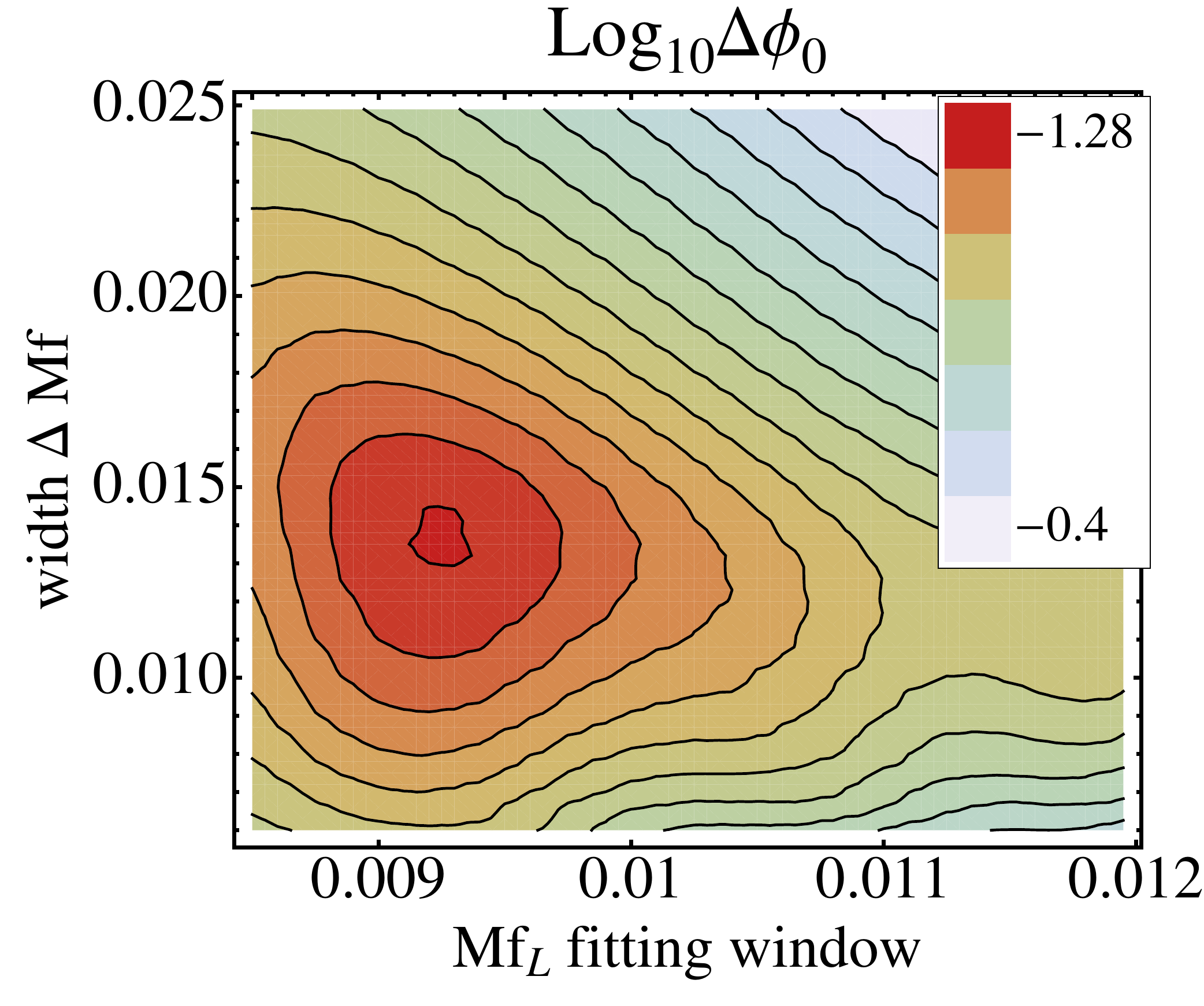} ~ \includegraphics[width=0.3\textwidth]{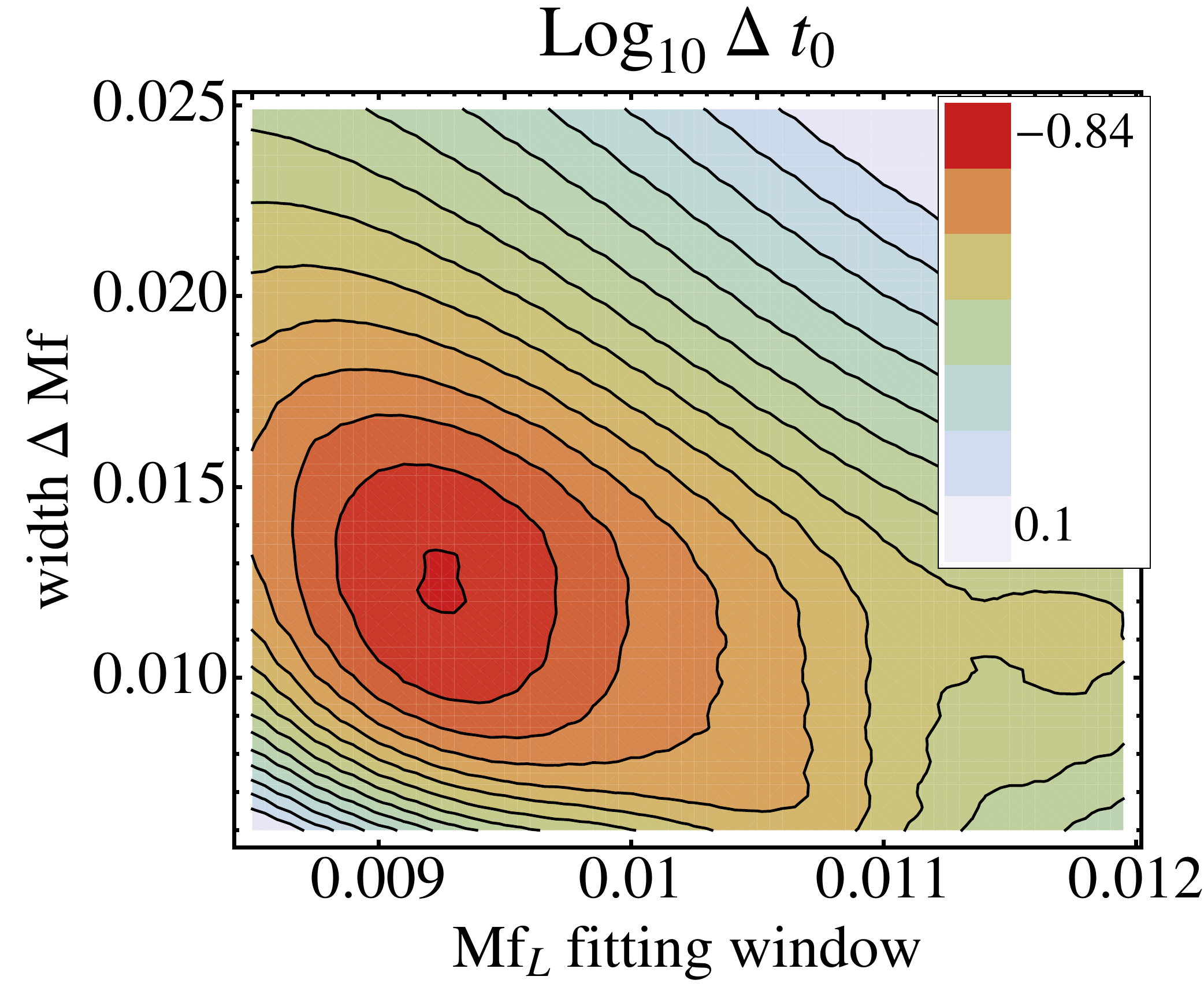}
  \caption{Dependence of the fitting errors in $\eta$, $\phi_0$ and $t_0$ on the frequency
    window $(f_L,\Delta f)$.  Note that there is a clear choice of
    $(f_L,\Delta f) \approx (0.0093,0.014)$ that optimizes the fit between the NR waveform
    and the PN waveforms with different $\eta$. For a binary of total
    mass $10\,M_\odot$, for which the last stable orbit happens at
    440~Hz, this corresponds to frequencies $(f_L,\Delta
    f)|_{10\,M_\odot}\approx(189,284)\,{\rm 
      Hz}$. This indicates that the optimal window for matching should
    start at the lowest reliable frequency available from the NR waveform and
    extend roughly up to the last stable orbit, which usually quantifies the
    point when the PN approximation starts to break down. Moreover,
    the plot on the left shows that the accuracy in 
    $\eta$ decreases slowly with different choices of $(f_L,\Delta
    f)$, assuring that small changes in these values do not lead to
    large errors in the hybrid construction.
    At the best fit
    point, the accuracy in $\eta$ by this fitting procedure is better
    than $10^{-3}$.}
  \label{fig:etaMerr}
\end{figure*}

Let us now consider the choice of the optimal matching window
$(f_L,f_L + \Delta f)$, and the best fit values of $(\phi_0,t_0,\eta_{\rm PN})$.
For each window, the least squares
procedure gives a best fit value $\eta_{\rm PN} = \eta(f_L,\Delta f)$ and
1-$\sigma$ error estimates $\Delta\eta, \Delta \phi_0, \Delta t_0$.
Our principle for choosing $(f_L,\Delta f)$ is to pick the one for
which the quality of fit between the NR and PN waveforms is the best,
i.e. to minimize the fitting errors.  

We first fix $\eta_{\rm PN}=
\eta_{\rm NR}$, choosing the 1:2 waveform, and consider
fitting for $(\phi_0,t_0)$. The result for $\Delta\phi_0$ is shown in
Fig.~\ref{fig:phaseError} as a contour plot in the $(f_L,\Delta f)$
plane.  There are clearly multiple best-fit islands but we already see
that the optimal window choice turns out to be a long frequency
width starting at low frequencies, or a relatively short window
starting closer to the merger. Regarding the increasing error PN most likely introduces
towards higher frequencies, we prefer using an early and long matching window.  
 Though we do not show it here, the
result is similar for the time offset $t_0$.  

\begin{figure}
  \includegraphics[width=0.85\columnwidth]{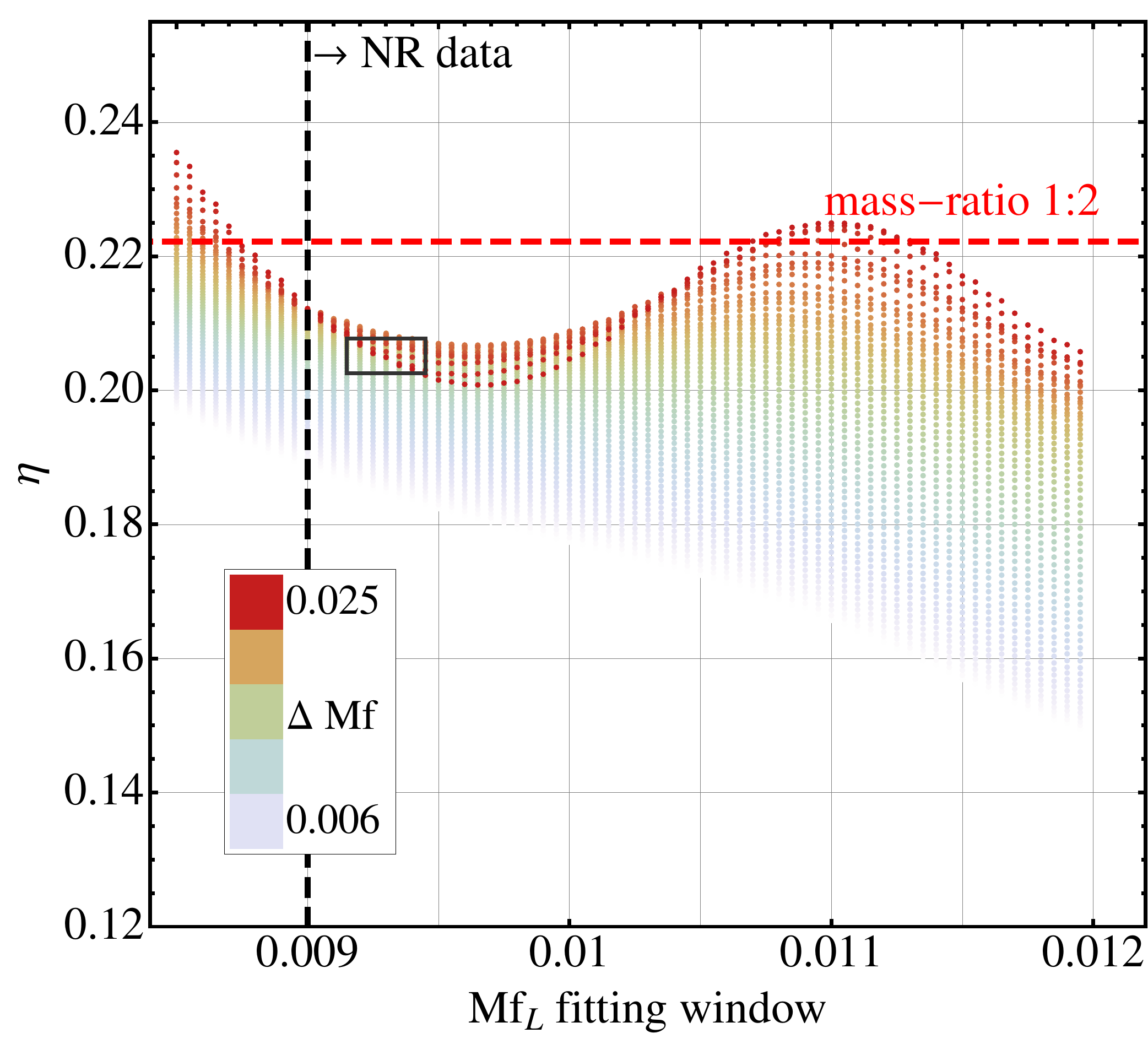}
  \caption{Best fit value of $\eta$ as a function of the start
    frequency $f_L$ of the matching window for the waveform which
    corresponds nominally to a mass ratio 1:2, i.e. $\eta_{\rm NR} = 2/9 =
    0.222\ldots$; this is shown by a horizontal dashed line. The
    vertical dashed line at $Mf_L = 0.009$ is the start frequency of
    the NR waveform. A rectangle highlights the region of minimal fitting 
    errors from Fig.~\ref{fig:etaMerr}.
   We see that the best determined values of $\eta$ are
    clearly less than $\eta_{\rm NR}$. }
  \label{fig:etaVal}
\end{figure}

It is more interesting
instead to generalize this and allow all three parameters
$(\eta_{\rm PN},\phi_0,t_0)$ to vary.  The main result is displayed in
Fig.~\ref{fig:etaMerr}, which shows contour plots of the fitting errors
$\Delta\eta$, $\Delta \phi_0$ and $\Delta t_0$ in the $(f_L,\Delta f)$ plane. 
There are now clear and consistent minima for all errors and thus a
clear best choice for $f_L$ and 
$\Delta f$.  At this optimal choice, we see that we can fit $\eta$,
$\phi_0$ and $t_0$ to 
better than $10^{-3}$, $0.06$ and $0.15 M$, respectively.
Apart from the error $\Delta\eta$, the actual best fit value $\eta$
is also of great interest.  Fig. \ref{fig:etaVal} shows the 
value of $\eta$ as a function of the start frequency of the
matching window $f_L$ and $\Delta f$.  The x-axis on this plot is the
start point of the fitting window $f_L$, and the color bar indicates
$\Delta f$.  The most trustworthy values correspond to the optimal choice of
$(f_L,\Delta f)$ obtained in Fig.~\ref{fig:etaMerr}; we indicate the
union of all three minimal-error islands as a rectangle in
Fig.~\ref{fig:etaVal}. 

To summarize, from Figs.~\ref{fig:etaMerr} and \ref{fig:etaVal} we
deduce that, if we were to ignore $\eta_{\rm NR}$ (the value that the
numerical simulation nominally assumes) and simply try to find the
best fit with the PN waveforms described in Sec.~\ref{sec:postnewton},
then we can clearly estimate the best matching region $(f_L,f_L +
\Delta f)$ and a best fit value $\eta_{\rm PN} = \eta\pm\Delta\eta$.
This procedure illustrates a trade-off between trying to match at
early frequencies, where our PN model is more reliable and having a
sufficiently long fitting window, in which a considerable frequency
evolution leads to an accurate estimate of the fitting parameters.
The difference between $\eta_{\rm NR}$ and $\eta_{\rm PN}$ for this case is
seen to be $\sim 10\%$. This by itself does not say that the
uncertainty in $\eta$ is $10\%$ because as we shall soon see, the
uncertainties in the hybrid waveform are dominated by the
uncertainties in the PN model. In other words, the NR waveform is
closer to the true physical waveform within the matching window, and
we should not actually use the best fit value of $\eta_{\rm PN}$ to
construct the hybrid.

\subsubsection{Accuracy of the hybrid waveform}

Later we shall show a phenomenological fit for the hybrid waveform and
we shall claim that the fit reproduces the hybrid waveform
sufficiently accurately. Here we first ask whether the hybrid
waveform is itself sufficiently accurate subject to various errors.
The basic criteria for evaluating this is the notion of a distance
between two signals whose difference is $\delta h$, as given in
Eq.~\eqref{eq:11}. For two signals $h$ and $h^\prime$, we shall
consider the normalized distance squared $(\delta h|\delta h)/\rho^2$, 
where $\rho$ is calculated from our best model (3PN amplitude, 3.5PN
TaylorF2 phase combined 
with highest resolution NR waveform). Now the total mass $M$ becomes
important.  Previously, when we 
looked at the least square fits in Eq.~\eqref{eq:9}, the total mass
appeared just as a scale factor.  However, in the inner product
Eq.~\eqref{eq:1}, the power spectral density $S_n(f)$ sets a frequency scale, and the
value for $(\delta h|\delta h)$ becomes mass-dependent.  We shall
consider two design noise curves, Initial and
Advanced LIGO~\cite{initialLIGO, advLIGO}. We are then addressing the
question of how different our  
hybrids would be if we were to use a slightly different result on either the NR or PN side.

\begin{figure*}
  \includegraphics[width=0.31\textwidth]{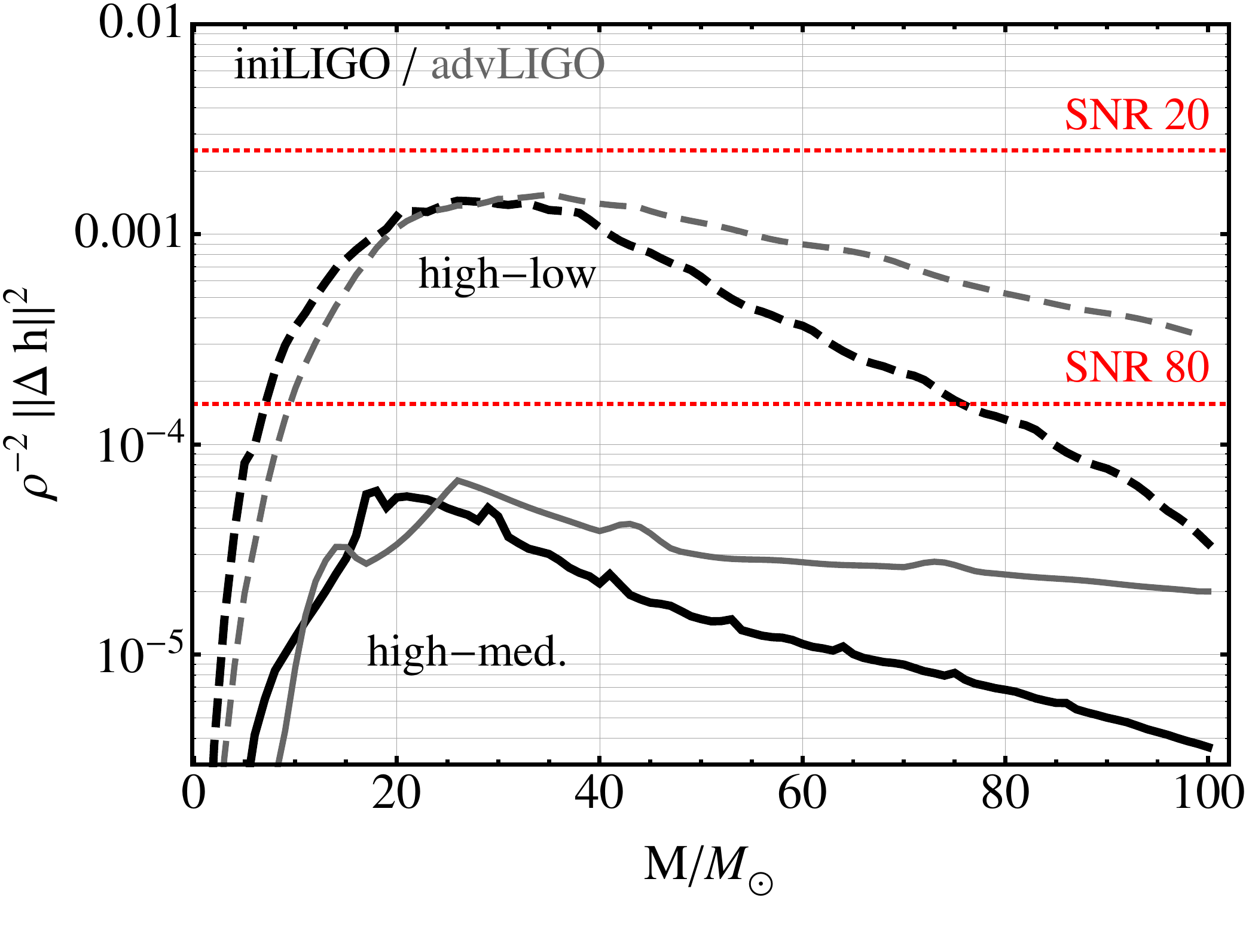}
  ~   \includegraphics[width=0.32\textwidth]{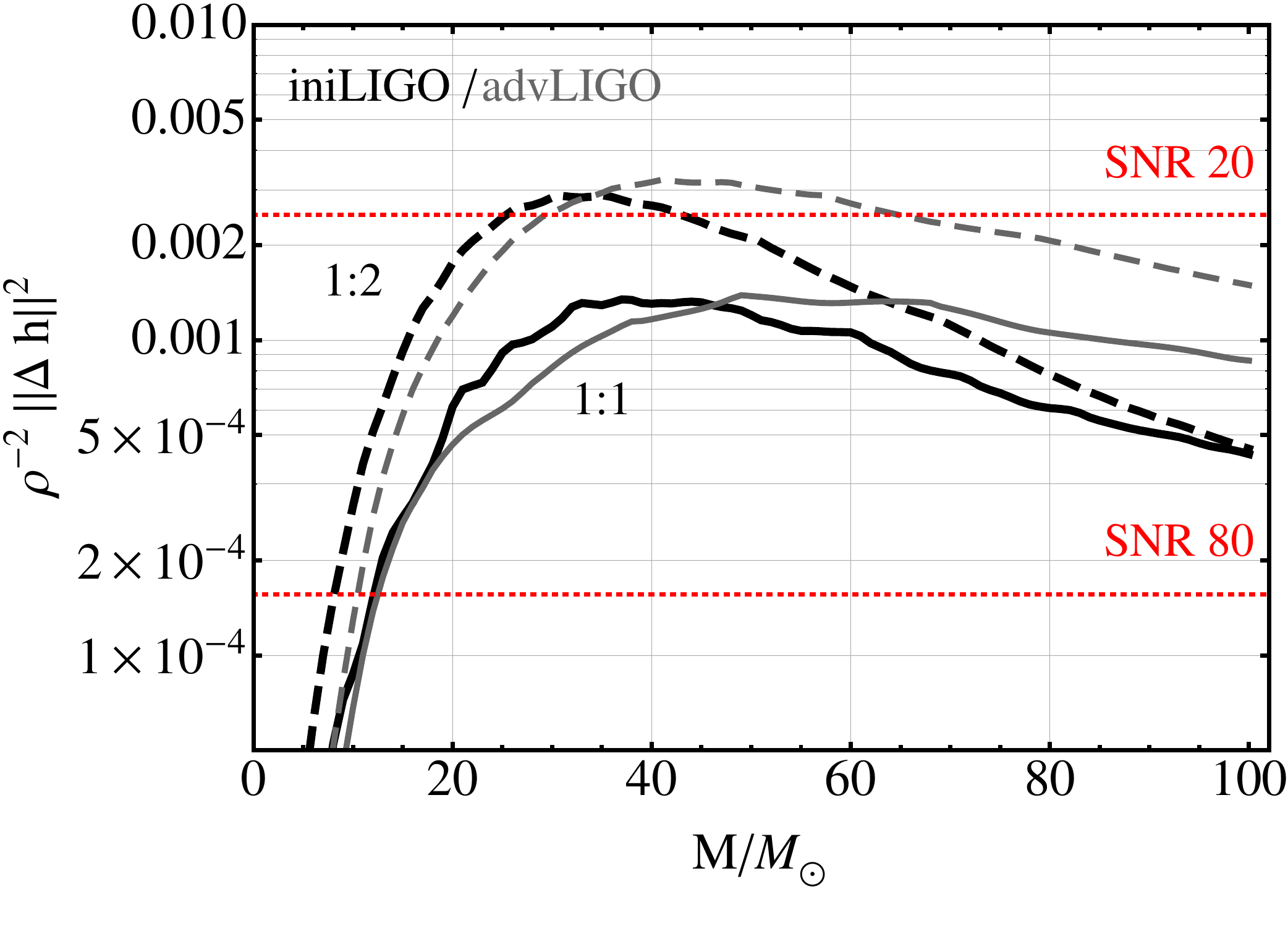} ~
  \includegraphics[width=0.32\textwidth]{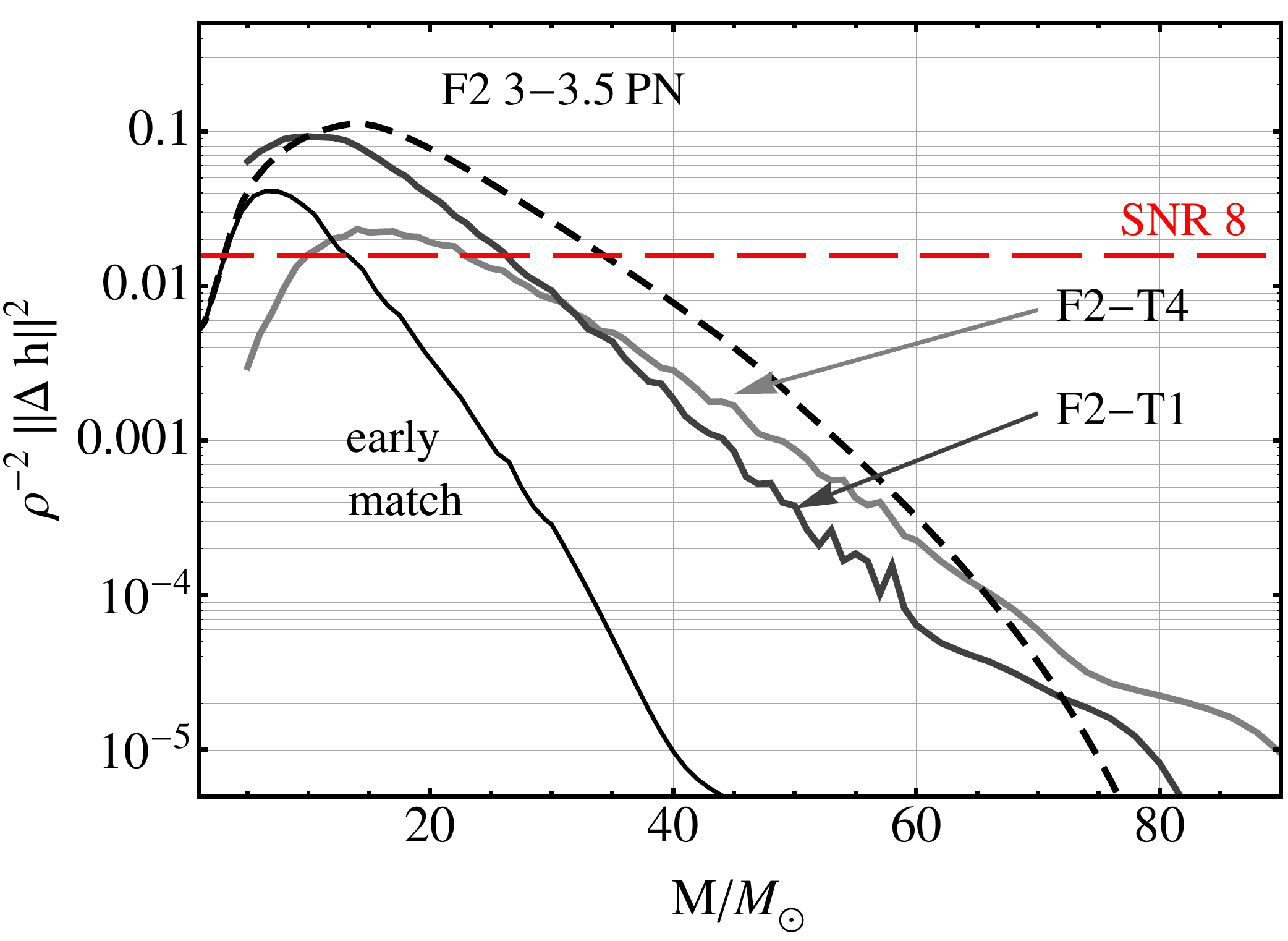}  

  \caption{Distinguishability of hybrid waveforms that have been
    constructed varying some of the hybrid ingredients at a time. When
    indicated, the black/grey color code denotes that Initial/Advanced
    LIGO design curves have been used for the distance
    calculation. The horizontal lines are the 
    lines of constant SNR (in fact it is $1/{\rm SNR}^2$); if the distance
    measure goes above them, then the waveforms can be
    distinguished from each other.
    The left panel shows the effect of constructing hybrids
    from \texttt{Llama} equal-mass waveforms at 
    different resolutions.  We consider
    the difference between the high-medium resolution waveforms, and
    the high-low waveform resolutions.  The central panel shows the
    effect of using NR waveforms produced with either \texttt{BAM} or 
    \texttt{Llama} codes. The solid lines indicate the normalized
    distance in the  
    equal-mass case, dashed lines show the case of mass-ratio 1:2. The 
    highest available resolution was always used. The panel on the
    right displays Initial LIGO's ability to distinguish hybrid waveforms
    constructed from different PN approximants.  This plot shows that
    the hybrids are not sufficient for detection at the
    $\epsilon=0.03$ level $[$Eq.~\eqref{eq:14}$]$ only for a small
    range of masses. ``Early match'' is a reference for matching 3PN
    or 3.5PN F2 at early frequencies to the long equal mass
    \texttt{SpEC} waveform.
\label{fig:DiffHybs}
}
\end{figure*}

On the NR side, we first consider data computed at different
resolutions. The \texttt{Llama} waveforms for the equal-mass case have
been computed at low, medium and high resolutions corresponding to
spacing h = 0.96, 0.80 and 0.64 on the wave extraction grid. The
finest grid, i.e. the grid covering the black hole, has a resolution
of $0.02$ for the finest resolution. This is scaled by $0.80/0.64$ and
$0.96/0.64$ for the medium and low resolution runs respectively.  We
combine these waveforms with the TaylorF2 model from
Sec.~\ref{sec:postnewton} by using the optimal matching window
discussed around Fig.~\ref{fig:phaseError} and $\eta_{\rm PN} =
\eta_{\rm NR}$.  The result is shown in the left panel of
Fig.~\ref{fig:DiffHybs}.  Hybrids constructed with medium- and
high-resolution waveforms would be indistinguishable even with
Advanced LIGO at a SNR of 80 over the considered mass range.  Thus, we
conclude that the numerical errors related to a finite resolution are
not relevant in the hybrid construction process.

The uncertainties increase when comparing NR 
data produced by different codes. Similar to the analysis of different
resolutions we calculate the distance of hybrid waveforms for
non-spinning black holes 
with mass ratio 1:1 and 1:2. Results from data set \#1 and \#8 (see
Table~\ref{tab:NRWaveforms}) 
were used, and the distance plot in the central panel of
Fig.~\ref{fig:DiffHybs} shows that the 1:2 waveform 
would be distinguishable for Advanced LIGO at SNR 20 for a total
masses between $\sim 30M_\odot$  and $\sim 65M_\odot$. 
Note that these errors are dominated by our matching to PN which
possibly yields different fit parameters 
for the PN model and therefore amplifies small differences in the NR
data. Towards higher masses, the  
influence of this matching decreases as well as the distance of both waveform.
However, as we shall show next, all these errors are still small compared 
to the intrinsic uncertainties introduced by PN and they do not matter
for Initial LIGO. 
If we care only about detection with a minimal match $\epsilon =
0.03$ $[$see Eq.~\eqref{eq:14}$]$, we have even less to worry about.  

The errors on the PN side turn out to be much more important.
The right panel of Fig.~\ref{fig:DiffHybs} illustrates the effect of
using different PN 
approximants combined with the same \texttt{SpEC} equal mass simulation.
We first use the fitting window discussed above, although
the exceptionally long \texttt{SpEC} waveform would allow a much
earlier matching. 
The dashed curve shows the difference in the hybrid waveforms when we
match the 3PN or 3.5PN phase following the TaylorF2 frequency domain
approximants described in Sec.~\ref{sec:postnewton} (the amplitude is
taken at 3PN order in both cases). 
 We see that the difference between
these hybrids becomes significant even for Initial LIGO at
SNR of 8 between a total mass of $\sim 5M_\odot$ and $\sim 35M_\odot$.
Similarly, the differences between the F2 and Taylor T1 \& T4
approximants are also significant.  For detection with $\epsilon =
0.03$ $[$see Eq.~\eqref{eq:14}$]$, we need to look at the horizontal line
with $(\delta h|\delta h)/\rho^2 = 0.06$ in
the right panel of Fig.~\ref{fig:DiffHybs}. Both in the 3PN/3.5PN
distance (dashed line) and the  
TaylorT1/TaylorF2 comparison (upper black solid line), there is a small range of
masses for which the difference between the hybrids would matter even
for detection.

As a reference, we make use of the fact that the numerical data \#9
(Table \ref{tab:NRWaveforms}) 
 actually contains physical information to frequencies considerably smaller 
than the matching window used for our hybrid production. We therefore
match the TaylorF2 phase at 3PN and 3.5PN  
order also with a much earlier fitting window (roughly a factor of $2$
lower in frequency). The right panel of Fig.~\ref{fig:DiffHybs} shows that 
the difference indicated as ``early match'' remains undetectable for a
larger range of total masses. Expanding such studies may be used to
quantify the necessary length of NR waveforms and estimate to which frequency
standard PN results can be used in hybrid waveform constructions.

Having carried out this study of errors for non-spinning waveforms, we
can now draw some conclusions for the aligned-spin case.  In
principle, the procedure outlined here remains valid; we should search
over not only $\{\eta,t_0,\phi_0\}$, but now also over the spins
$\{\chi_1,\chi_2\}$.  We would not expect the results to be
\emph{better} than shown here for non-spinning waveforms because (i)
we are adding two more parameters and (ii) the waveforms \#1-4 are
expected to have more wave-extraction systematic errors than the
\texttt{Llama} results considered here.  Most importantly, as we have
just seen, the intrinsic errors in PN are more significant whereas the
numerical accuracy is not the bottleneck. The intrinsic parameter
biases in PN also show up when different PN models are compared with
each other.  An extensive comparison of different PN models is made in
\cite{Buonanno:2009zt}; this paper quantifies the mutual effectualness
and faithfulness of the different PN models and shows that errors of
$\sim 20\%$ are not uncommon for Advanced LIGO.  The less than 10\%
discrepancy in $\eta$ shown in Fig.~\ref{fig:etaVal} is thus entirely
consistent with the differences between different PN models.  To
address this, one needs either improved PN models or a greater variety
of longer NR waveforms such as the long \texttt{SpEC} simulation.

As a simplification, in what follows below we will choose the matching
window based on maximizing over the extrinsic parameters
$(t_0,\phi_0)$ motivated by Fig.~\ref{fig:phaseError}.  In that
figure, we observe the best fit region extending diagonally from
$M\Delta f \approx 0.013$ on the y-axis, to the bottom right corner.
It turns out that for this diagonal, the upper frequency of the window
does not vary much, $0.020 \lesssim Mf_L + M\Delta f \lesssim 0.024$,
 and we shall use this fact
below for constructing hybrid waveforms for aligned spinning systems.

\subsection{Construction of hybrid waveforms for aligned spinning
  systems} 
\label{sec:hybridConstruction}

Let us now proceed to the construction of a hybrid waveform model for
non-precessing, spinning systems with comparable mass.  Again, the
waveforms described in Sec.~\ref{sec:postnewton} will be the basis for
our model at low frequencies corresponding to the inspiral stage. On
the other hand, the NR simulations described as data-sets \#1--3 in
Table~\ref{tab:NRWaveforms} contain physical information for
frequencies above $Mf \approx 0.009$.  We will refer to
Fig.~\ref{fig:phaseError} to justify our choice of an overlapping
window at $Mf \in (0.01,0.02)$.  

Once this interval is fixed, we now carry out the
following matching procedure for all NR
simulations of data-sets \#1--3: PN and NR phases are aligned by
fitting the free parameters $t_0$ and $\phi_0$ in Eq.~(\ref{eq:F2phase}); 
with a standard root-finding algorithm (starting at the mid point
of the fitting interval) we find a frequency $f_\Phi$ where PN and NR phase 
coincide and construct the hybrid phase
consisting of TaylorF2 at $f \leq f_\Phi$ and NR data at $f > f_\Phi$.
An analogous procedure is applied to the amplitude, but 
in this case there is no freedom for adjusting any parameters. Hence,
we use an educated
guess for the matching frequency (compatible with that for the
phase) and find the root $f_A$ where the difference of PN and NR
amplitude vanishes. The hybrid amplitude consists of
PN data before and NR data after $f_{A}$. Small
wiggles in the NR amplitude, due to the Fourier transform, 
do not affect the phenomenological fit significantly. The most
important ingredient for arriving at an effectual model is the phase. 

\begin{figure*}
\includegraphics[width=0.43\textwidth]{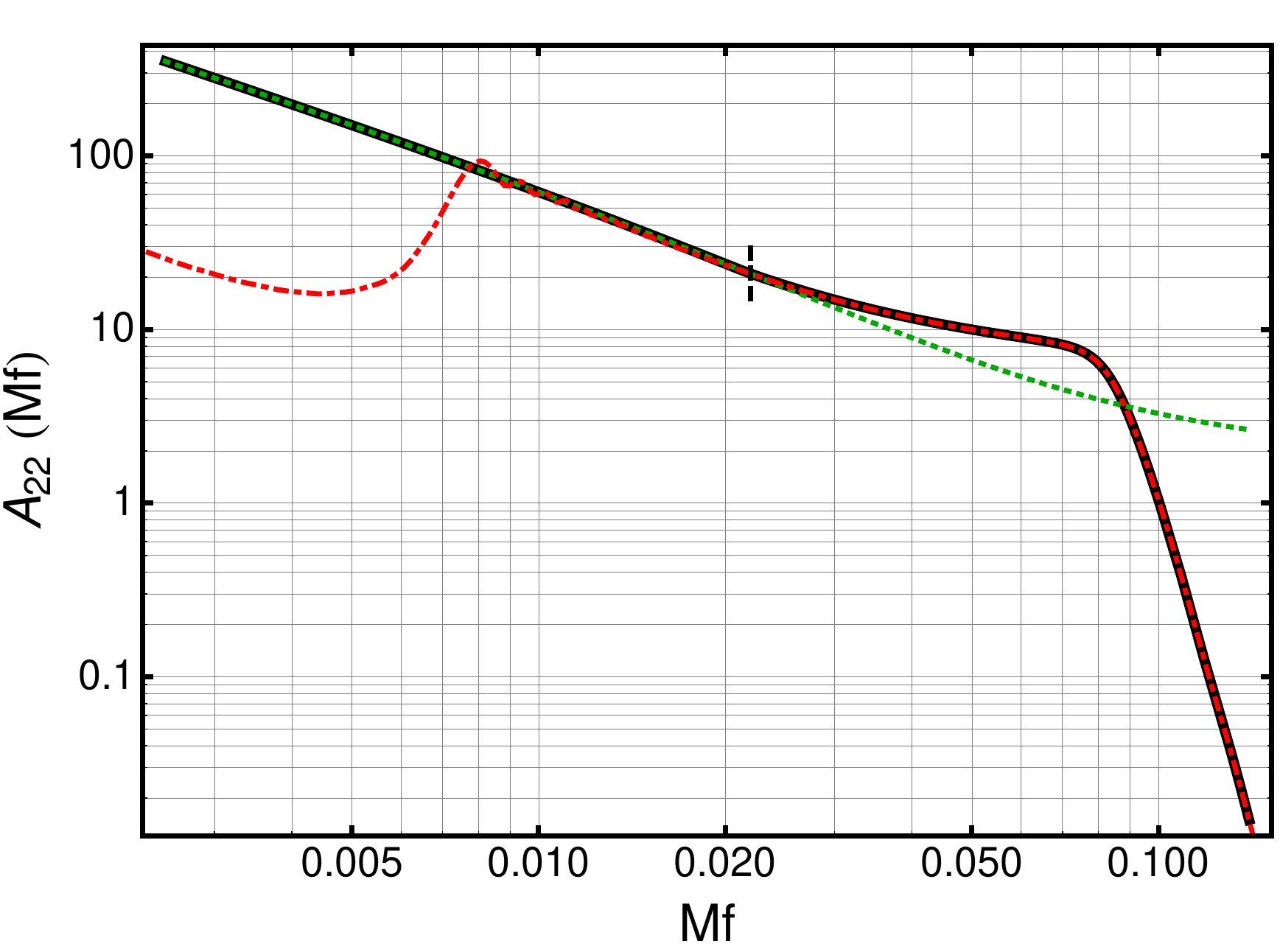} ~
\includegraphics[width=0.43\textwidth]{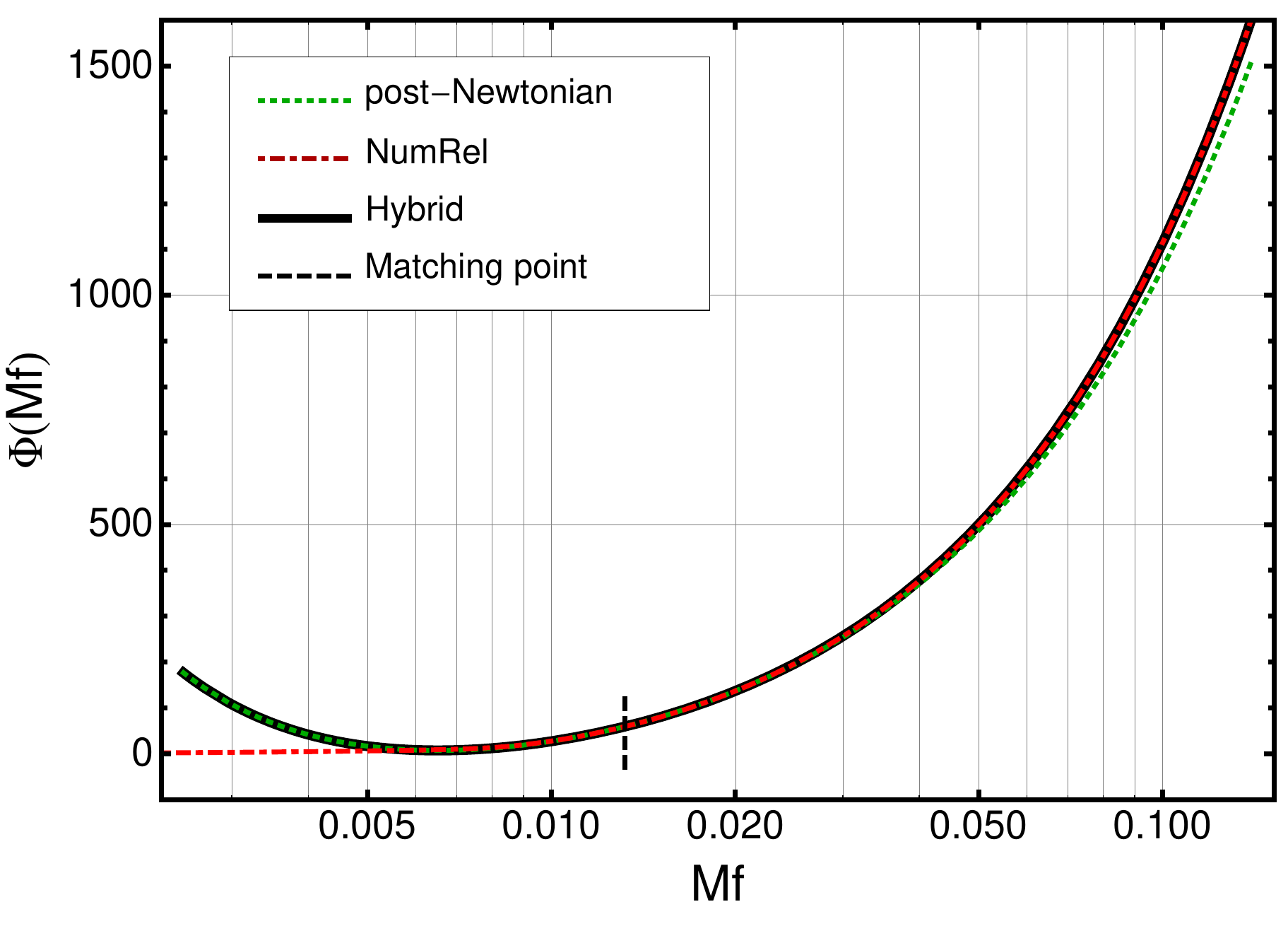}
\caption{Illustration of the method for constructing PN-NR hybrid
    waveforms in the frequency domain. The data corresponds to an
    equal-mass binary with aligned spins $\chi_1 = \chi_2 = -0.25$. The
    left panel shows the amplitude and the right panel displays the phase
    of the dominant $\ell = 2, m = 2$ mode of the GW complex strain
    $ \tilde{h}(f)$. The
    green dotted lines correspond to the TaylorF2 PN approximant and
    the red dot-dashed curve is the NR data. The hybrid
    waveform is depicted in solid black and the matching points for
    amplitude and
    phase are indicated with a dashed line.
\label{fig:hybridConstr}}
\end{figure*}

Fig.~\ref{fig:hybridConstr} illustrates the above-described hybrid
construction method for 
matching PN and NR data in the frequency domain. The procedure
does not require any re-sizing the 
PN or NR data and allows for the construction of waveforms
containing all the information from the TaylorF2 approximant at low
frequencies and input from the NR simulations for the late
inspiral, merger and ringdown. 
The resulting hybrid PN-NR data cover a part of the 
parameter space corresponding to equal-valued, (anti-)aligned spins for
$0.16 \leq \eta \leq 0.25$ and constitute the \lq\lq target\rq\rq 
waveforms to be fitted by the analytical phenomenological model described in
Section~\ref{sec:fitting}. 

\section{Phenomenological model}
\label{sec:fitting}

In this section we present the phenomenological model developed in
order to fit the hybrid PN-NR waveforms of Section~\ref{sec:pn-vs-nr}
to an analytical formula. A geometric description of the procedure for
constructing phenomenological waveforms parametrized by just the
physical parameters is detailed in \cite{Ajith:2007kx}, and here we
just summarize it.  Let $\mathcal{M}$ be the space of intrinsic
physical parameters that we are interested in. In the present case,
this is the four-dimensional space of the component masses and spins
$\lambda = \{M,\eta,\chi_1,\chi_2\}$.  For each point $\lambda$ in
$\mathcal{M}$, let $h(t;\lambda)$ be the true physical waveform that
we wish to approximate; in particular we consider only the dominant
$\ell=m=2$ mode in this paper.  Furthermore, as in
\cite{Ajith:2009bn}, we model the spin effects using a single
parameter $\chi$ defined as
\begin{equation}
  \label{eq:chidef}
  \chi \equiv \frac{1 + \delta}{2} \chi_1 + \frac{1-\delta}{2} \chi_2,   
\end{equation}
where $\delta \equiv (m_1-m_2)/M$. As mentioned in
Sec.~\ref{sec:numrel}, this is justified from PN treatments of the
inspiral~\cite{Vaishnav:2007nm} and from numerical simulations of the
merger~\cite{Reisswig:2009vc} which considers equal mass systems.  In
these works, it is found that the dominant spin effect on the waveform
is from the \emph{mass-weighted total spin} of the system.  On the PN side,
this can
be further justified by looking at the expressions for the PN phase
and amplitudes given in the appendix
[Eqs.~(\ref{eq:F2Coeffs},\ref{eq:ampCoeffs})].  Consider the phase and
amplitude terms as polynomials in $\eta$ and retain only the
$\mathcal{O}(\eta^0)$ terms.  At this lowest order, the amplitude and
phase are seen to depend only on $\chi$. We can therefore hope that
this single parameter captures the main effects of the black hole
spins, at least for the purpose of constructing an effectual model.
In fact, whenever we incorporate pure PN contributions in our final model, we
use them with $\chi_1 = \chi_2 = \chi$.
It is however important to note that this is only an approximation;
while it suffices for our purposes (i.e., in constructing an effectual
model) it will need to be refined as more faithful models are
required.  The degeneracy in the space of aligned spins will be
studied in greater detail in a forthcoming paper
\cite{Ohme-Proc}, but here we look to construct a
phenomenological model using only $(M,\eta,\chi)$ as the physical
parameters.

We start with some known signals in this parameter space at $N$ points
$\lambda_1,\lambda_2,\ldots,\lambda_N$.  We take these known signals
to be the hybrid waveforms whose construction we described earlier.
Here the NR waveforms are the \texttt{BAM} waveforms of data sets
\#1-3 summarized in Table~\ref{tab:NRWaveforms}, and the PN model is
the 3.5PN frequency domain model for aligned spins described in
Sec.~\ref{sec:postnewton}.  Given the finite set of hybrid waveforms
constructed from these ingredients, we wish to propose a
phenomenological model $h_{\rm phen}(t;\lambda)$ that interpolates
between the hybrid waveforms with sufficient accuracy.  In
constructing this phenomenological model, it is convenient to work not
with the physical parameters $\lambda$, but rather with a larger set
of phenomenological parameters $\tilde{\lambda}$, which we shall
shortly describe.  If $\widetilde{\mathcal{M}}$ is the space of
phenomenological parameters, then we need to find a one-to-one mapping
$\mathcal{M} \rightarrow \widetilde{\mathcal{M}}$ denoted
$\tilde{\lambda}(\lambda)$, and thus the subspace of
$\widetilde{\mathcal{M}}$ corresponding to the physical parameters. As
the end result of this construction, for every physical parameter
$\lambda$, we will know the corresponding phenomenological parameter
$\tilde{\lambda}(\lambda)$ and thus the corresponding phenomenological
waveform $h_{\rm phen}(t;\tilde{\lambda}(\lambda))$.

Following the construction procedure of
Section~\ref{sec:hybridConstruction}, we split our waveforms in
amplitude and phase, both of which shall be fitted to a
phenomenological model
\begin{equation}
\tilde{h}_{\rm phen}(f) = A_{\rm phen}(f) \, e^{i \Phi_{\rm phen}(f)}.
\label{eq:phenomAnsatz}
\end{equation}
For both the amplitude and the phase of the dominant mode of the GW
radiation, we make use of the insights from PN and perturbation
theory for the description of the inspiral and ringdown
of the BBH coalescence, respectively, and introduce a phenomenological
model to complete the description of the waveforms in the merger.

\subsection{Phase model}
\label{subsec:phase}

The PN approach for the GW radiation based on the stationary phase
approximation, introduced in Eq.~\eqref{eq:F2phase} of
Section~\ref{sec:postnewton} (used with
$t_0 = \phi_0 = 0$ and $\chi_1 = \chi_2 = \chi$), gives an adequate
representation of the phase of the dominant mode during the adiabatic inspiral
stage of the BBH coalescence $\psi^{22}_{\rm SPA}(f)$.  
As the system transitions towards the merger phase, it is expected
that further terms in the expansion are required to capture the
features of the evolution. With this ansatz in mind, we propose a
pre-merger phase $\psi^{22}_{\rm PM}(f)$ of the form
\begin{align}
\psi^{22}_{\rm PM}(f) &=\frac{1}{\eta} \left( \alpha_1 f^{-5/3} +
\alpha_2 f^{-1}\right. \notag \\ &+ \alpha_3 f^{-1/3}   \left. +
\alpha_4 + \alpha_5 f^{2/3} + \alpha_6 f\right), 
\label{eq:preMergPhase}
\end{align}
where the $\alpha_k$ coefficients are inspired by the SPA phase,
redefined and phenomenologically fitted to
agree with the hybrid waveforms in the region between the frequencies
$0.1 f_{\rm RD}$ and $f_{\rm RD}$, which depend on the spins and
masses of the black holes in the form explained below in
Eq.~\eqref{eqn:Cardoso_fofMa}.  Note that $0.1 f_{\rm RD}$ roughly
corresponds to the starting frequency of our NR simulations.

As for the post-merger phase, the Teukolsky
equation~\cite{Teukolsky:1973ha} describes the ringdown
of a slightly distorted spinning black hole. The metric perturbation for the
fundamental mode at 
large distances can be expressed as an exponential damped sinusoidal
\begin{equation}
h_{\rm ring}^{22}(t) = \frac{\mathcal{A}_{\rm ring}M}{D_L} \,
e^{-\pi f_{\rm RD} t/Q} \, e^{- 2\pi i f_{\rm RD}\, t}, \label{eq:damped_RD}
\end{equation}
where $M$ is the mass of the ringing black hole, $D_L$ the distance
from the source, and $Q$ and $f_{\rm RD}$ correspond to the quality factor
of the ringing down and the central frequency
of the quasi-normal mode. These can be
approximated with an error $\leq 2.5\%$ in the range
$a \in [0,0.99]$ by the following fit~\cite{Berti:2005ys}
\begin{align}
f_{\rm RD}(a,M)&=\frac{1}{2 \pi} \frac{c^3}{GM}\left[ k_1 + k_2
  (1-a)^{k_3}\right],
\label{eqn:Cardoso_fofMa}
\\ 
Q(a)&=q_1 + q_2(1-a)^{q_3}, 
\label{eqn:Cardoso_Qofa}
\end{align}
where $k_i = \{1.5251, -1.1568, 0.1292 \}$ and $q_i = \{ 0.7000,$
$1.4187, -0.4990 \}$, $i=1,2,3$ as given in Table VIII
of~\cite{Berti:2005ys} for the $(l,m,n)=(220)$ mode. The
review~\cite{Berti:2009kk} presents a full description of 
quasi-normal modes. 
The quantity $aM^2$ is the spin magnitude of the final black hole
after the binary has 
merged, which can be inferred from the spins of the two black
holes. In our case, we use the fit presented
in~\cite{Rezzolla:2007rd}, which maps the mass-ratio and spins of the
binary to the total spin $a$ of the final black hole. 

\begin{figure}
  \includegraphics[width=\columnwidth]{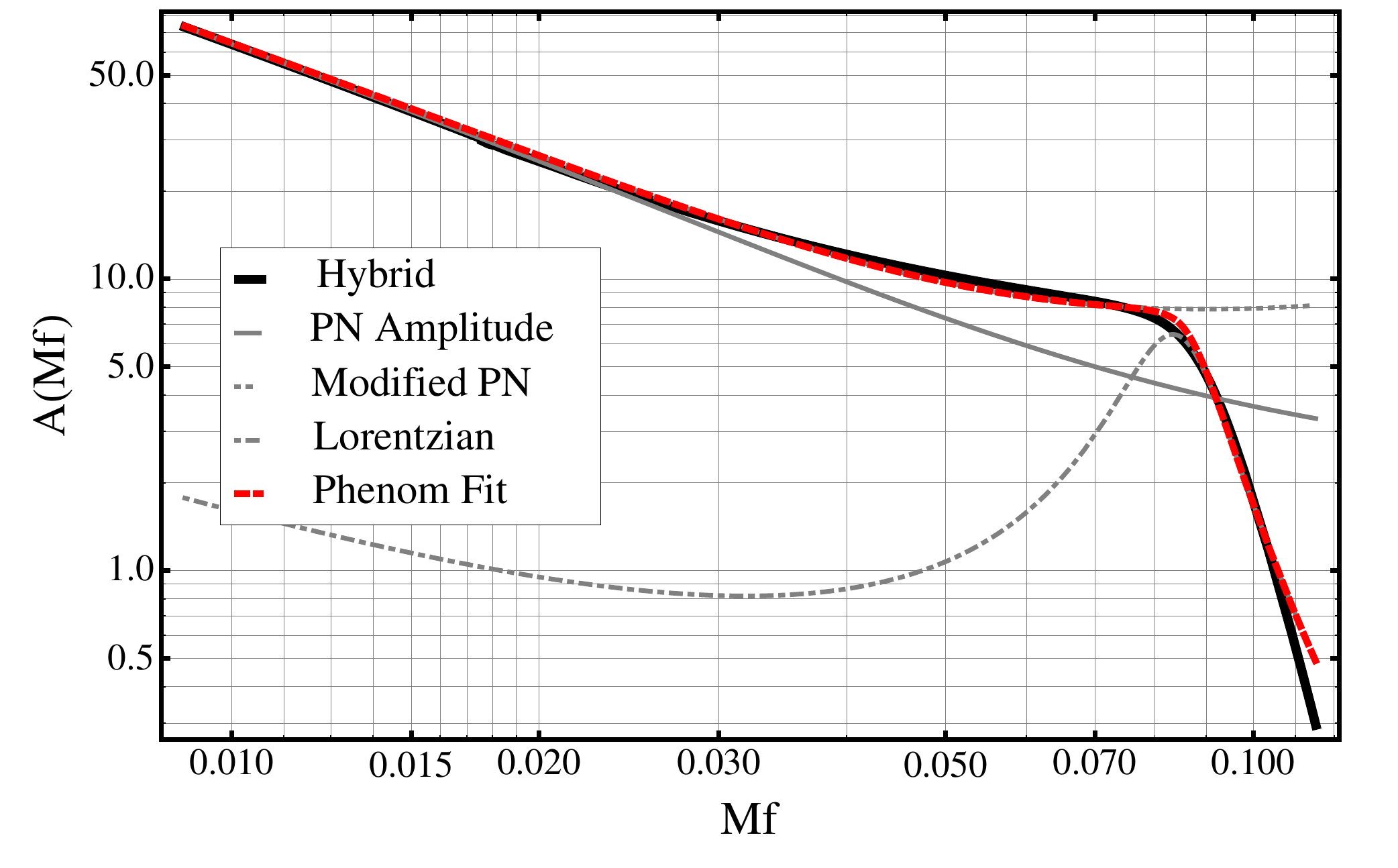}
  \caption{Fitting procedure for the amplitude, applied to the
    equal-mass, non-spinning case. The $\gamma_1$
    term of Eq.~\eqref{eq:preMergAmpl} is introduced to follow
    the behavior of the amplitude in 
    the pre-merger regime whereas the Lorentzian curve correctly
    describes the post-merger. The two pieces are glued together in a
    smooth manner using tanh-windows.}
  \label{fig:fitAmplitude}
\end{figure}
\label{subsec:ampl}

The analytical treatment of the ringdown (\ref{eq:damped_RD})
motivates a linear ansatz for the post-merger 
phase $\psi^{22}_{\rm RD}(f)$ of the form
\begin{equation}
\psi^{22}_{\rm RD}(f) = \beta_1 + \beta_2 f.
\label{eq:RDPhase}
\end{equation}
The $\beta_{1,2}$ parameters are not fitted, but obtained from the
pre-merger ansatz~\eqref{eq:preMergPhase} by taking the value and slope
of the phase at the transition point $f_{\rm RD}$.
The transition between the different regimes is smoothened by means of
tanh-window functions 
\begin{equation}
 w^{\pm}_{f_0} =
\frac{1}{2}\left[1\pm\tanh\left(\frac{4(f-f_0)}{d}\right)\right]
\end{equation}
to 
produce the final phenomenological phase
\begin{equation}
\Phi_{\rm phen}(f)=\psi^{22}_{\rm SPA}w^{-}_{f_1}
+ \psi^{22}_{\rm PM} w^+_{f_1}w^-_{f_2}
+\psi^{22}_{\rm RD}w^{+}_{f_2},
\label{eq:smoothPhase}
\end{equation}
with $f_1=0.1 f_{\rm RD}$, $f_2=f_{\rm RD}$; here we have used $d=0.005$
in the window functions $w^\pm$. Roughly, these two transition
points respectively signal the frequencies at which our NR simulations
start and the point at which the binary merges, and have been found to
provide the best match between the hybrids and the phenomenological
model.

\subsection{Amplitude model}

In a similar manner to the phase, we approach the problem of fitting
the amplitude of the GW by noting that the PN amplitude obtained
from the SPA expression could be formally re-expanded as
\begin{equation}
\tilde{A}_{\rm PN}^{\rm exp}(f) = C \Omega^{-7/6} \left( 1 + 
\sum_{k=2}^5 \gamma_k \Omega^{k/3} \right),
\label{eq:SPAAmpl}
\end{equation}
where $\Omega = \pi M f$. We introduce a higher-order term to model the
pre-merger amplitude $\tilde{A}_{\rm PM}(f)$
\begin{equation}
\tilde{A}_{\rm PM}(f) = \tilde{A}_{\rm PN}(f) +
\gamma_1 f^{5/6},
\label{eq:preMergAmpl}
\end{equation}
where the $\gamma_1$ coefficient is introduced to model the amplitude
in the pre-merger regime and $\tilde A_{\rm PN}$ is the amplitude constructed
in Sec.~\ref{sec:postnewton} (see Fig.~\ref{fig:PNamps}).

The ansatz for the amplitude during the ringdown is
\begin{equation}
\tilde{A}_{\rm RD}(f) = \delta_1 {\cal  L}\left(f,f_{\rm
    RD}(a,M),\delta_2Q(a)\right) f^{-7/6},   
\label{eq:RDAmpl}
\end{equation}
where only the width and overall magnitude of the Lorentzian function
${\cal L}(f,f_0,\sigma) \equiv \sigma^2/\left( (f-f_0)^2 +\sigma^2/4
\right)$ are fitted to the hybrid data. The factor $f^{-7/6}$ is
introduced to correct the Lorentzian at high frequencies, since the
hybrid data shows a faster fall-off, and $\delta_1$ accounts for the
overall amplitude scale of the ringdown.  In principle, the
phenomenological parameter $\delta_{2}$ should not be necessary
because the width of the Lorentzian for the ringdown should be given
by the quality factor $Q$ which depends only on the spin of the final
black hole.  However, recall that here we estimate the final spin from
the initial configuration using the fit given
in~\cite{Rezzolla:2007rd}; $\delta_2$ accounts for the errors in this
fit.  

The phenomenological amplitude is constructed from these two pieces in
a manner analogous to the phase
\begin{equation}
\tilde{A}_{\rm phen}(f)=\tilde{A}_{\rm PM}(f)
w^{-}_{f_0}+\tilde{A}_{\rm RD}(f)w^{+}_{f_0},
\label{eq:smoothAmp}
\end{equation}
with $f_0 = 0.98 f_{\rm RD}$ and
$d=0.015$. Fig.~\ref{fig:fitAmplitude} demonstrates how this
phenomenological ansatz fits the hybrid amplitude in a smooth
manner through the late inspiral, merger and ringdown. 

\subsection{Mapping the phenomenological coefficients}

\begin{figure*}
  \includegraphics[width=0.9\textwidth]{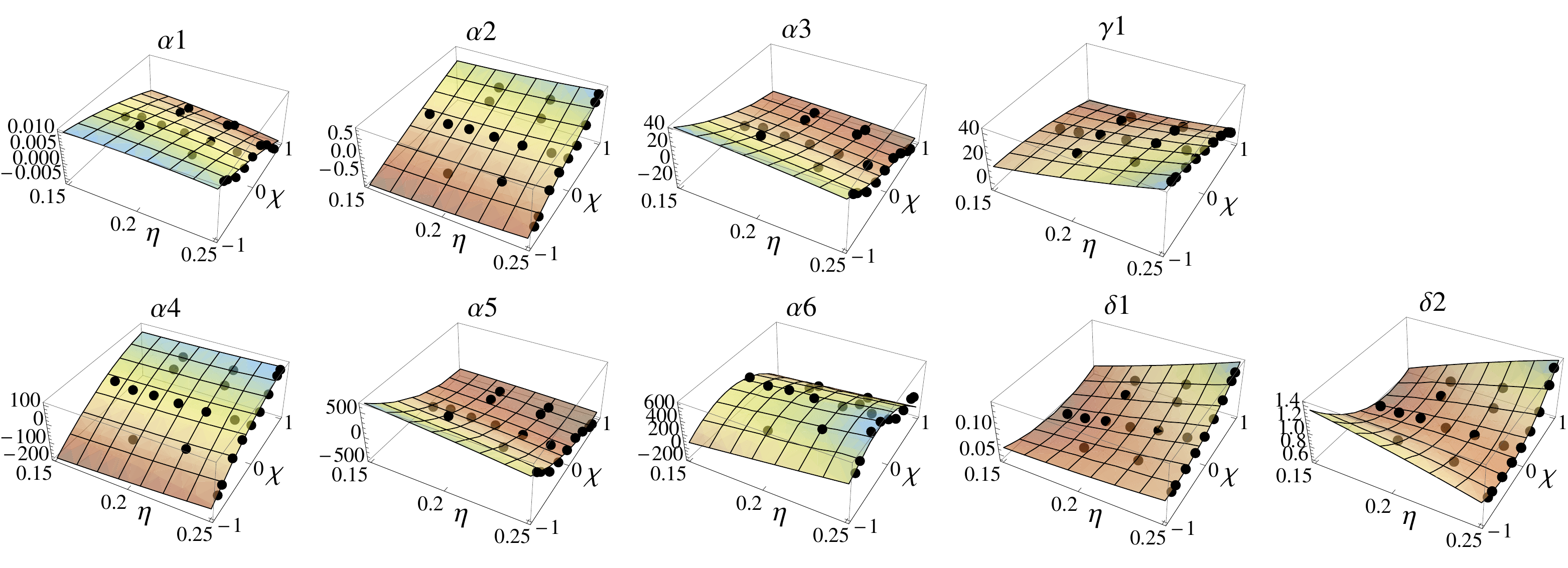}
  \caption{Map of the phenomenological parameters to the physical
    parameters of the binary.}
  \label{fig:mapToPhysPars}
\end{figure*}

\begin{table*}[t]
\caption{Coefficients to map the 9 free parameters of our 
phenomenological model to the physical parameters of the binary.} 
\label{tab:map}
\begin{center}
\begin{tabular*}{0.96\linewidth}{@{\extracolsep{\fill}}c | c c c c c c }
\hline
\hline
$\Lambda_k$ & $\zeta^{(01)}$ & $\zeta^{(02)}$ & $\zeta^{(11)}$ &
$\zeta^{(10)}$ & $\zeta^{(20)}$ \\ 
\hline

$\alpha_1$ & $-2.417\times10^{-3}$ & $-1.093\times10^{-3}$ & $-1.917\times10^{-2}$ & $7.267\times10^{-2}$ & $-2.504\times10^{-1}$ \\

$\alpha_2$ & $5.962\times10^{-1}$ & $-5.6\times10^{-2}$ & $1.52\times10^{-1}$ & $-2.97$ & $1.312\times10^{1}$ \\

$\alpha_3$ & $-3.283\times10^{1}$ & $8.859$ & $2.931\times10^{1}$ & $7.954\times10^{1}$ & $-4.349\times10^{2}$ \\

$\alpha_4$ & $1.619\times10^{2}$ & $-4.702\times10^{1}$ & $-1.751\times10^{2}$ & $-3.225\times10^{2}$ & $1.587\times10^{3}$ \\

$\alpha_5$ & $-6.32\times10^{2}$ & $2.463\times10^{2}$ & $1.048\times10^{3}$ & $3.355\times10^{2}$ & $-5.115\times10^{3}$ \\

$\alpha_6$ & $-4.809\times10^{1}$ & $-3.643\times10^{2}$ & $-5.215\times10^{2}$ & $1.87\times10^{3}$ & $7.354\times10^{2}$ \\

\hline

$\gamma_1$ & $4.149$ & $-4.07$ & $-8.752\times10^{1}$ & $-4.897\times10^{1}$ & $6.665\times10^{2}$ \\

\hline

$\delta_1$ & $-5.472\times10^{-2}$ & $2.094\times10^{-2}$ & $3.554\times10^{-1}$ & $1.151\times10^{-1}$ & $9.64\times10^{-1}$ \\

$\delta_2$ & $-1.235$ & $3.423\times10^{-1}$ & $6.062$ & $5.949$ & $-1.069\times10^{1}$ \\

\hline
\hline
\end{tabular*}
\end{center}
\end{table*}

Our models for the amplitude and phase involve 9 phenomenological
parameters $\{\alpha_1,\ldots,\alpha_6, \gamma_1,\delta_1, \delta_2\}$
defined in Eqs.~\eqref{eq:preMergPhase}, \eqref{eq:preMergAmpl} and
\eqref{eq:RDAmpl}.  The coefficients $\beta_{1,2}$
from~\eqref{eq:RDPhase} can be trivially derived from the set of
$\alpha_k$. We now need to find the mapping $\mathcal{M}\rightarrow
\widetilde{\mathcal{M}}$ from the physical to these phenomenological
parameters. As mentioned earlier, instead of $\chi_{1,2}$, we consider
$\chi$, the weighted sum of the spins, defined in Eq.~\eqref{eq:chidef}.
Thus, our phenomenological waveforms are parametrized only by the
symmetric mass ratio $\eta$ and the spin parameter $\chi$, as well as
by the total mass of the system $M$ through a trivial
rescaling. Fig.~\ref{fig:mapToPhysPars} shows the mapping of
$\alpha_k$, $\gamma_k$ and $\delta_k$ to surfaces in the
$(\eta,\chi)$--plane.

The 9 phenomenological coefficients introduced in our model, denoted
generically by $\Lambda_k$, are 
expressed in terms of the physical parameters of the binary as
\begin{equation}
\Lambda_k = \sum_{i+j \in \{1,2\}} \zeta^{(ij)}_k \eta^i \chi^j,
\label{eq:map}
\end{equation}
which yields 5 coefficients $\zeta^{(ij)}$ for each of the 9
parameters, as given in table~\ref{tab:map}.  

\begin{figure}[t]
  \includegraphics[width=0.5\textwidth]{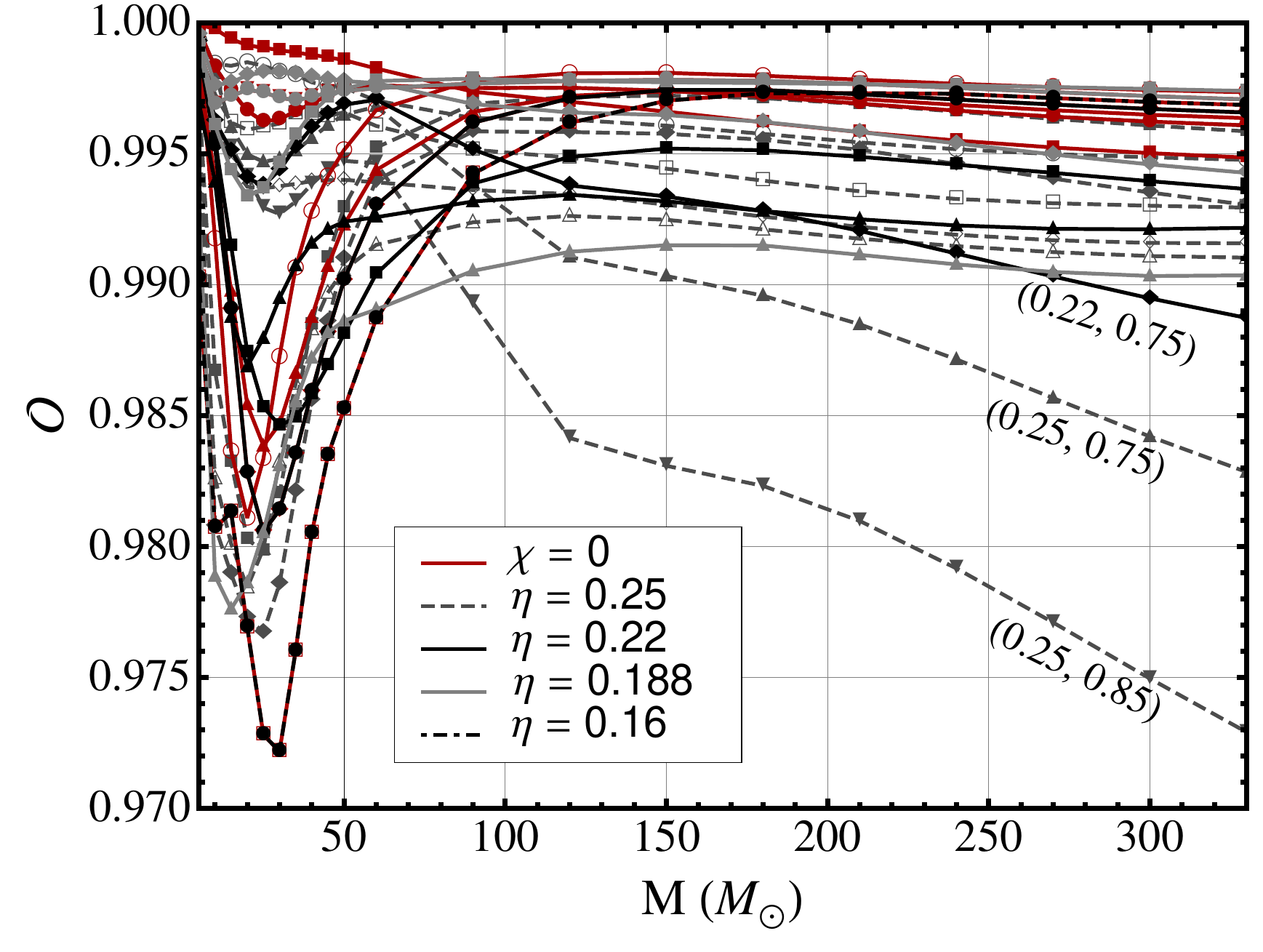}
  \includegraphics[width=0.5\textwidth]{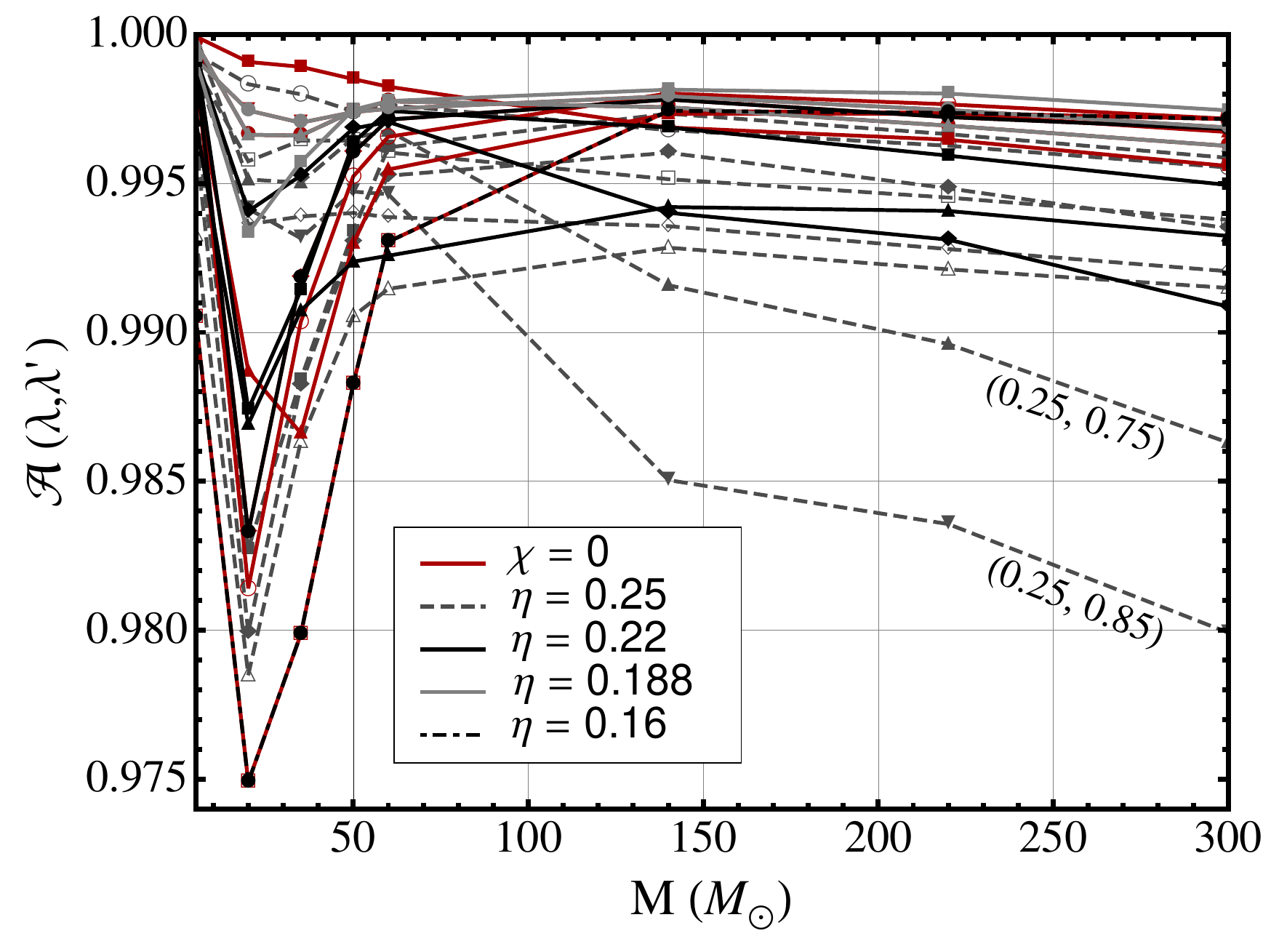}
  \caption{Overlaps and fitting factors between the hybrid waveform
    constructed according 
    to the procedure described in Sec.~\ref{sec:hybridConstruction}
    and the proposed phenomenological fit, using the design
    sensitivity curve of Advanced LIGO. The
    labels indicate the values of ($\eta,\chi$) for some
    configurations. In the upper panel We plot
    $\mathcal{O}(\lambda) = \mathcal{A}(\lambda,\lambda)$, i.e. we
    compute the ambiguity function \eqref{eq:12} without
    maximizing over the parameters of the model waveform; this is a
    lower bound on the effectualness. The bottom panel shows the
    maximized overlaps, i.e. $\mathcal{A}(\lambda,\lambda^\prime)$;
    the maximum bias of the optimized $\lambda^\prime$ parameters is
    $\Delta\eta=5\times 10^{-3},\Delta\chi=5\times10^{-2},\Delta M =
    3\,M_\odot$. 
  \label{fig:Overlaps}
}
\end{figure}

We evaluate the goodness of fit between the phenomenological model and
the hybrid waveforms in terms of the fitting factor, i.e. the ambiguity
function $\mathcal{A}(\lambda,\lambda^\prime)$ defined in
Eq.~\eqref{eq:12} and the overlap, i.e. $\mathcal{O}=
\mathcal{A}(\lambda,\lambda)$.  In evaluating the overlap, we maximize over the
extrinsic parameters $t_0,\phi_0$ as indicated in Eq.~\eqref{eq:12},
but for the results shown in the upper panel of
Fig.~\ref{fig:Overlaps} we do not perform the additional maximization over
the model parameters $\lambda^\prime$. Thus, the results shown there 
can be viewed as a lower bound on the effectualness. The maximization
over the intrinsic parameters $\eta,\chi$ and $M$ allows to study the
faithfulness of the model.

\begin{figure}[t]
  \includegraphics[width=0.5\textwidth]{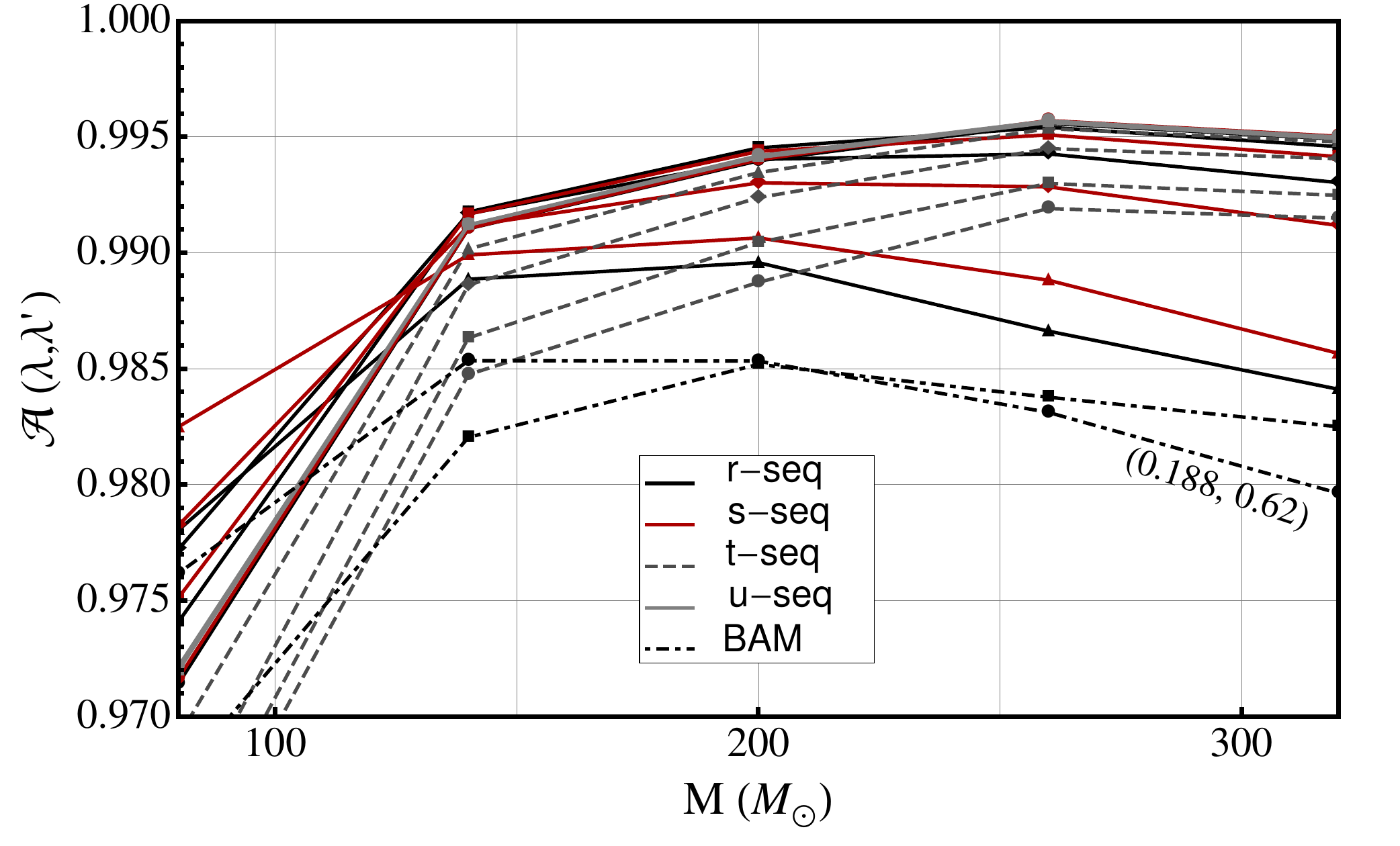}\\
  \caption{Upper panel: maximized overlaps
    $\mathcal{A}(\lambda,\lambda^\prime)$ between 
    the NR data-sets \#4--7ab and the 
    predicted phenomenological waveforms from our model for Advanced
    LIGO.  The
    labels indicate the values of ($\eta,\chi$) for some
    configurations. Note that the 
    short duration of the NR data prevents us from computing overlaps
    at lower masses. The maximization in this case has been done over
    $\eta$ and $\chi$ keeping $M$ fixed and the maximum bias on the maximized
    parameters is $\Delta 
    \eta = 6\times10^{-3}, \Delta{\chi} =
    5\times10^{-2}$. 
  \label{fig:OverlapsCompr}
}

\end{figure}

Figs.~\ref{fig:Overlaps} and~\ref{fig:OverlapsCompr} illustrate the result
using the design curve of the Advanced LIGO detector. Fig.~\ref{fig:Overlaps}
shows the overlap and fitting factor between the hybrid waveforms
constructed in  
Sec.~\ref{sec:hybridConstruction} and 
their corresponding phenomenological fit. The match 
approaches unity by construction at low masses and degrades with
increasing total mass. Nevertheless, for none of the hybrid waveforms
employed in 
the construction of our model does the overlap fall below a
value of $\sim 0.97$, thus reflecting the fact that the phenomenological
model effectually represents the target signals. A further maximization over
the $\lambda^\prime$ parameters, shown in the lower panel of
Fig.~\ref{fig:Overlaps}, indicates a
maximum bias on the intrinsic parameters of the binary of
$\Delta\eta=5\times 10^{-3},\Delta\chi=5\times10^{-2},\Delta M = 
3\,M_\odot$.

We have constructed a gravitational waveform model for binary black
hole inspiral and coalescence starting with a particular set of
simulations, and using a particular ansatz for the waveform.  Is this
model robust, and is it consistent with waveforms from other numerical
simulations?  In the upper panel of Fig.~\ref{fig:OverlapsCompr}, and
as a further test to assess the robustness of our model, we compute
the maximized overlap between the phenomenological waveforms and the NR
data-sets \#4-7 that were {\em not} used in the construction of the
model. At low masses, there is no contribution of these short NR
waveforms in the frequency band of interest for Advanced LIGO, and it
turns out that the overlaps can be computed only for $M\gtrsim
100M_\odot$.

In Fig.~\ref{fig:OverlapsCompr} we see that
the maximization of the overlaps with respect to $\eta$ and $\chi$
shows values $>0.97$ for all configurations; in this case
the maximum bias in the parameters is $\Delta 
    \eta \approx 6\times10^{-3}, \Delta{\chi} \approx
    5\times10^{-2}$. This is roughly consistent with
Fig.~\ref{fig:Overlaps} which shows the overlap and fitting factor of
the model with the 
original set of hybrid waveforms.  
These results prove that our model is
effectual and, thus, sufficient for detection. We shall study its
effectualness and faithfulness in greater detail in a forthcoming
paper.


\section{Summary and future work}
\label{sec:summary}

The aim of this paper has been to construct an analytical model for
the inspiral and coalescence of binary black hole systems with aligned
spins and comparable masses in circular orbits.  Since this requires
merging post-Newtonian and numerical relativity waveforms, one of the
main themes has been to quantify the internal consistency of hybrid
waveforms. This is important because even if one succeeds in finding a
useful fit for a family of hybrid waveforms, one still needs to show
that the hybrid one started with is a sufficiently good approximation
to the true physical waveforms.  We investigated the systematics of
constructing hybrid waveforms for accurate non-spinning waveforms
based on the \texttt{Llama} code and we saw that neither the numerical errors 
nor the hybrid-construction errors are significant.  This suggests
that in order  
to improve the accuracy of hybrid waveforms, we require either longer 
NR waveforms so that the matching with PN can be done earlier in the 
inspiral phase, or improved PN models that are more accurate at frequencies
closer to the binary merger.

With the hybrid waveforms for non-precessing systems in hand, we
constructed an analytical model for the waveform which has an overlap
and fitting factor of better than 97\% for Advanced LIGO with the
hybrid waveforms for systems with a total mass ranging up to $\sim 350
M_\odot$. Since these overlaps are comparable to those achieved with
the alternative phenomenological waveform construction presented
in~\cite{Ajith:2009bn}, we conclude that this process is robust, and,
in particular, its accuracy is not affected by the way in which the
transitions between inspiral, merger and ringdown are modeled.
Furthermore, though we have not discussed it in detail in this paper,
it turns out that the model presented here agrees very well with the
model of~\cite{Ajith:2009bn}. This will be discussed in detail in a
forthcoming paper \cite{Ohme-Proc}

In the future we will study in greater detail the effectualness and
faithfulness of this waveform model, thereby quantifying more
precisely its performance for detection and parameter estimation.  In
this context it is important to extend this work to modes higher than
the dominant $\ell = 2$, $m = \pm2$ spherical harmonics.  It was shown
recently \cite{McWilliams:2010eq} that the overlap with the real
signal can possibly be affected by the inclusion of higher modes up to
the order of $\sim 1\%$, which is comparable or greater than the
disagreement we find between hybrid and phenomenological model.

We will further quantify the behavior of our templates in real
non-Gaussian detector noise, and use them in real searches for
gravitational wave signals. Eventually, work is underway in extending
the model to include precessing spins. Our phenomenological model can
be readily applied to existent GW detection efforts within the
LIGO/Virgo Scientific Collaborations. Ongoing searches are already
making use of inspiral-merger-ringdown waveforms, such as the EOBNR
family and the phenomenological family
of~\cite{Ajith:2007qp,Ajith:2007kx,Ajith:2007xh,Ajith:2009bn} in the
form of software injections and as filter approximants. Our newly
developed frequency-domain matching procedure should serve to
cross-check the validity of these alternative approaches and to
complement them.

\section*{Acknowledgments}
We thank Doreen M\"uller for carrying out some of the {\tt BAM}
simulations, and Stas Babak, Vitor Cardoso, Steve Fairhurst, Ian
Hinder, Doreen M\"uller, Dirk P\"utzfeld, Bangalore Sathyaprakash and
Bernard Schutz 
for useful comments and discussions. LS has been partially supported by
DAAD grant A/06/12630. MH was supported by FWF Lise-Meitner project M1178-N16 at
the University of Vienna. SH was supported by DAAD grant D/07/13385 and
grant FPA-2007-60220 from the Spanish Ministry of Science.
DP has been supported by grant CSD-2007-00042 of the Spanish Ministry
of Science. DP and CR received support from the Bundesministerium f\"ur
Bildung und Forschung, Germany.
{\tt BAM} simulations were performed at computer centers
LRZ Munich, ICHEC Dublin, VSC Vienna, CESGA Santiago the Compostela and
at MareNostrum at Barcelona Supercomputing Center --
Centro Nacional de Supercomputaci\'on (Spanish National
Supercomputing Center).
This work was supported in part by the DFG grant SFB/Transregio 7
``Gravitational wave astronomy'' and by the DLR (Deutsches Zentrum f\"ur Luft- und 
Raumfahrttechnik).

\appendix

\begin{widetext}
\section{PN expansion coefficients}

For the convenience of the reader we explicitly give all the PN
expansion coefficients used in Section~\ref{sec:postnewton} as
functions of the symmetric mass ratio
$\eta$~(\ref{eq:symmmassratio}), the dimensionless spin magnitudes \mbox{$\chi_i = (\bm{S_i} \cdot \bm{\hat
    L})/m_i^2$}, where $\bm{\hat L}$ is the unit angular momentum
vector, and $\chi = \chi_1 \, m_1/M  + \chi_2 \, m_2/M $.  
The energy~(\ref{eq:PNenergy}) is given in terms of
\begin{align}
 e_0 &= 1 , 
\qquad e_1 = 0 , 
\qquad e_2 = -\frac{3}{4} -\frac{\eta }{12} ,  \qquad
 e_3 =  \frac{8}{3} \chi - \frac23 \eta (\chi_1 +
 \chi_2) ,   \nonumber \\
e_4 &= -\frac{27}{8} +  \frac{19\eta}{8} 
-\frac{\eta ^2}{24} - \chi^2  ,   \label{eq:energyCoeff}\\
e_5 &= \frac{72 - 31 \eta}{9} \chi 
 - \frac{45 \eta - \eta^2}{9}  (\chi_1+\chi_2)  ,   \nonumber\\
e_6 &= -\frac{675}{64} + \eta\left(\frac{34445}{576}-\frac{205 \pi
    ^2}{96}\right) -\frac{155 \eta ^2}{96} -\frac{35 \eta ^3}{5184} .  \nonumber
\end{align}%
The flux coefficients read
\begin{align} 
 f_0 &= 1 ,\qquad
 f_1 = 0 , \qquad
f_2 =-\frac{1247}{336}-\frac{35 \eta }{12} , \qquad
f_3 = 4 \pi - \frac{11}{4} \chi + \frac{3 \eta}{2}
(\chi_1 + \chi_2 )  , \nonumber \\ 
f_4 &= -\frac{44711}{9072} + 2 \chi^2 + \eta \left(\frac{9271}{504} - \frac{\chi_1\chi_2}{8}  \right) +
 \frac{65 \eta ^2}{18}  ,   \nonumber \\ 
 f_5 &= -\pi \left( \frac{8191}{672} + \frac{583 }{24} \eta \right) 
- \chi \left( \frac{63}{16} - \frac{355}{18} \eta \right)
   + (\chi_1 + \chi_2) \left(\frac{25}{8}\eta
  -\frac{157}{18} \eta^2 \right) - \frac{3}{4} \chi^3 + \frac{9 \eta}{4} \chi 
\,
\chi_1 \chi_2  ,   \label{eq:fluxCoeffs}\\ 
f_6 &= \frac{16 \pi ^2}{3} + \frac{6643739519}{69854400} - \frac{1712
  \gamma_E }{105}-\frac{856}{105} \ln \left(16 x\right)   + \eta
\left( \frac{41 \pi ^2}{48} -\frac{134543}{7776} 
 \right)-\frac{94403 \eta ^2}{3024} -\frac{775 \eta ^3}{324} ,  \nonumber\\ 
f_7 &= \pi \left( - \frac{16285}{504}+\frac{214745  \eta }{1728}+
  \frac{193385 \eta ^2}{3024} \right) . \nonumber
\end{align} 
$\gamma_E \approx 0.5772$ is the Euler constant. Note that the
next-to-leading order spin-orbit effects appearing at relative 2.5PN order
($f_5$) have recently been corrected \cite{PhysRevD.81.089901} and we take these
corrections into account.

The TaylorT4 approximant can be written as a series (\ref{eq:T4}) with
the following coefficients 
\begin{align} 
 a_0 &= 1 , \qquad a_1 = 0 , \qquad a_2 = -\frac{743}{336}-\frac{11
   \eta }{4} , \qquad
a_3 = 4 \pi - \frac{113}{12}  \chi + \frac{19 \eta}{6}  (\chi_1 + \chi_2 ) , \nonumber \\  
a_4 &= \frac{34103}{18144} + 5 \chi^2  + \eta \left( \frac{13661}{2016} - \frac{\chi_1 \chi_2}{8}
\right) + \frac{59 \eta ^2}{18} , \label{eq:T4Coeffs} \nonumber \\ 
a_5 &= - \pi \left(\frac{4159 }{672} + \frac{189}{8} \eta\right) -
\chi \left( \frac{31571}{1008} - \frac{1165}{24} \eta
\right)  + (\chi_1 + \chi_2) \left( \frac{21863}{1008} \eta - \frac{79}{6}
\eta^2\right) 
 - \frac{3}{4} \chi^3 + \frac{9 \eta}{4} \chi  \, \chi_1 \chi_2  ,    \\ 
\begin{split}
a_6 &= 
\frac {16447322263}{139708800} - \frac {1712}{105}\, \gamma_E +
\frac{16 \pi^{2}}{3}-\frac {856}{105} \ln  \left(16 x \right)  + \eta \left ( 
\frac {451 {\pi}^{2}}{48} - \frac {56198689}{217728} \right )
+{\frac {541}{896}}\,{\eta}^{2}
-{ \frac {5605}{2592}}\,{\eta}^{3}  \\
& \quad - \frac{80 \pi}{3} \chi + \left( \frac{20 \pi}{3} - \frac{1135}{36} \chi \right)\eta (\chi_1 + \chi_2)
+ \left( \frac{64153}{1008} - \frac{457 }{36} \eta  \right) \chi^2 - \left( \frac{787 }{144} \eta  - \frac{3037 }{144} \eta ^2 \right) \chi_1 \chi_2, 
\end{split}
   \nonumber \\
\begin{split}
 a_7 &= - \pi \left( \frac {4415}{4032} -\frac {358675}{6048} \eta
  -\frac {91495}{1512} \eta^{2} \right) 
- \chi \left( \frac{2529407}{27216} - \frac{845827 }{6048} \eta +
\frac{41551}{864} \eta ^2 \right) \\
& \quad + (\chi_1 + \chi_2) \left( \frac{1580239 }{54432} \eta
-\frac{451597}{6048}  \eta ^2 + \frac{2045}{432}  \eta ^3 
+\frac{107 \eta  }{6} \chi ^2 -\frac{5 \eta ^2}{24} \chi_1 \chi_2 \right) + 12
\pi \, \chi^2 \\
& \quad - \chi^3 \left(\frac{1505}{24} + \frac{\eta}{8} \right) + \chi \, \chi_1 \chi_2 \left( \frac{101  }{24} \eta + \frac{3}{8} \eta^2 \right)
. \end{split}
 \nonumber 
\end{align}
The spin-dependent terms that appear at 3 and 3.5 PN order (i.e., in $a_6$ and
$a_7$) are not complete since the
corresponding terms are not known in energy and flux. However, in this re-expansion they do appear as contributions from lower order
spin effects and we keep them.

The TaylorF2 description of the Fourier phase~(\ref{eq:F2phase}) is
expressed in terms of 
\begin{align}\nonumber
\alpha_0 &= 1, \qquad \alpha_1 = 0, \qquad \alpha_2 = \frac{3715}{756}
+ \frac{55 \eta}{9} ~, \qquad  
\alpha_3 = -16 \pi + \frac{113}{3} \chi - \frac{38 \eta}{3} (\chi_1 +
\chi_2)~, \\ \nonumber
\alpha_4 &= \frac{15293365}{508032} - 50 \chi^2
+ \eta \left(\frac{27145 }{504} + \frac{5}{4} \chi_1 \chi_2  \right)
+ \frac{3085 \eta ^2}{72} ~, \\ \label{eq:F2Coeffs}  
\begin{split}
\alpha_5 &= \left[ 1 + \ln \left( \pi f \right) \right] \left[ \pi \left(
    \frac{38645}{756} - \frac{65}{9} \eta \right) 
- \chi \left( \frac{735505}{2268} + \frac{130 }{9} \eta
\right) 
+ (\chi_1 + \chi_2) \left( \frac{12850}{81}  \eta +
  \frac{170}{9} \eta ^2\right) \right. \\ & \qquad \qquad \qquad
\left. - \frac{10}{3} \chi^3 + 10 \eta \chi \, \chi_1 \chi_2
 \right] , 
\end{split} 
\\ \nonumber 
\begin{split}
 \alpha_6 &= \frac{11583231236531}{4694215680} - \frac{640 \pi ^2}{3}
 - \frac{6848 }{21} \gamma_E - \frac{6848}{63} \ln \left(64 \pi
   f\right)  
 + \eta \left(   \frac{2255 \pi
     ^2}{12}-\frac{15737765635}{3048192}\right) 
 +\frac{76055 }{1728} \eta^2 \\ & \quad -\frac{127825 }{1296} \eta^3
+ \frac{2920 \pi}{3} \chi - \frac{175 - 1490 \eta}{3} \chi^2- \left(\frac{1120 \pi }{3} -\frac{1085 }{3} \chi \right) \eta (\chi_1 + \chi_2) 
+ \left( \frac{26945 }{336} \eta-\frac{2365 }{6} \eta ^2 \right) \chi_1 \chi_2
,  
\end{split} \\ \nonumber
\begin{split}
 \alpha_7 &= \pi \left( \frac{77096675}{254016}+\frac{378515  }{1512}
  \eta-\frac{74045 }{756} \eta ^2 \right) 
- \chi \left( \frac{20373952415}{3048192}  + \frac{150935 }{224}\eta 
-  \frac{578695 }{432}\eta ^2 \right) \\ & \quad
+ (\chi_1 + \chi_2) \left( \frac{4862041225}{1524096} \eta  +
\frac{1189775}{1008} \eta ^2 - \frac{71705}{216} \eta ^3
-\frac{830 \eta  }{3} \chi ^2 +\frac{35 \eta ^2 }{3} \chi_1 \chi_2\right) - 560
\pi \, \chi^2 \\ & \quad
+ 20 \pi \eta \, \chi_1 \chi_2 + \chi^3 \left(\frac{94555}{168}-85 \eta \right) + \chi \, \chi_1 \chi_2 \left(\frac{39665 }{168}\eta +  255 \eta ^2 \right)
. 
\end{split}
\end{align}
The comment we just made about the spin contributions at 3 and 3.5 PN order holds for the $\alpha$-coefficients of the TaylorF2 phase as well. Also, note that the contributions in $\alpha_5$ that are not proportional to $\ln (\pi
f)$ could be absorbed in a re-definition of the undetermined additional phase $\phi_0$ that appears in Eq.~(\ref{eq:F2phase}). 
(A similar discussion can be found in \cite{Arun:2004hn}.)
However, since we chose to set $\phi_0 = 0$ when combining this phase description with other analytical formulas in our 
phenomenological model (\ref{eq:smoothPhase}), it is important to keep all terms in $\alpha_5$.

The time-domain amplitude coefficients collected from
\cite{Blanchet:2008je,Berti:2007nw,Arun:2008kb} read 
 \begin{align} \nonumber
   \mathcal A_0 &= 1 , \qquad \mathcal A_1 = 0, \qquad \mathcal A_2 =
   -\frac{107}{42} + \frac{55}{42} \eta , \qquad 
\mathcal A_3 = 2 \pi - \frac{4}{3} \chi + \frac{2 \eta}{3} (\chi_1 + \chi_2) , \\  
\mathcal A_4 &= -\frac{2173}{1512} - \eta \left( \frac{1069  }{216} -
  2 \chi_1 \, \chi_2 \right) + \frac{2047 }{1512} \eta ^2 , \qquad 
\mathcal A_5 = -\frac{107 \pi }{21} + \eta \left( \frac{34 \pi }{21}
  -24 i \right) , \label{eq:ampCoeffs}\\ \nonumber 
\mathcal A_6 &=
\frac{27027409}{646800}-\frac{856\gamma_E}{105}+\frac{428 i \pi
}{105}+\frac{2 \pi ^2}{3} + \eta \left( \frac{41 \pi^2}{96}
  -\frac{278185}{33264}\right) -\frac{20261 \eta 
^2}{2772}+\frac{114635 \eta^3}{99792}-\frac{428}{105} \ln (16 x) .
 \end{align}

\end{widetext}

\bibliography{NewPhenom}

\end{document}